\documentclass[aps,prd,showpacs,notitlepage,nofootinbib,superscriptaddress,floatfix,showkeys,twocolumn]{revtex4-2}

\usepackage{blindtext}
\usepackage{hyperref}
\usepackage{amsmath,amssymb}
\usepackage{float}
\usepackage{microtype}
\usepackage{graphicx}
\usepackage{bm}
\usepackage{latexsym}
\usepackage{epsfig}
\usepackage{psfrag}
\usepackage{color}
\usepackage[dvipsnames]{xcolor}
\usepackage{subfigure}
\usepackage{placeins}
\usepackage{calligra}
\usepackage{mathrsfs}
\usepackage[scr=rsfs]{mathalpha}
\usepackage[normalem]{ulem}
\usepackage{bbm}

\newcommand{\sr}{r}

\hypersetup{
  colorlinks   = false, 
  urlcolor     = blue, 
  linkcolor    = blue, 
  citecolor   = blue 
}             
             
\newcommand{\cAs}{{\mathcal{A}_{\bmk}^{\sigma}}}
\newcommand{\cAsm}{{\mathcal{A}_{-\bmk}^{\sigma}}}
\newcommand{\cAsmp}{{\mathcal{A}_{-\bmk}^{\sigma}}{}\!\!\!\!'\,\,}

\newcommand{\Sch}{Schr\"{o}dinger }
\newcommand{\bmk}{{\bm k}}

\def\nn{\nonumber} 
\def\f{\frac}

\def\l{\left}
\def\r{\right}
\def\d{{\mathrm{d}}}
\def\pa{\partial}
\def\Mpl{M_{_{\mathrm{Pl}}}}
\def\cA{{\mathcal{A}}}
\def\cP{{\mathcal{P}}}

\def\pb{\mathcal{P}_{_{\mathrm{B}}}}
\def\pe{\mathcal{P}_{_{\mathrm{E}}}}

\def\HI{H_{\mathrm{I}}}

\def\ee{\eta_{\mathrm{e}}}
\def\e1i{\epsilon_{1\mathrm{i}}}
\def\Ne{N_{\mathrm{e}}}
\def\ke{k_{\mathrm{e}}}

\def\ns{n_{_{\mathrm{S}}}}
\def\phii{\phi_{\mathrm{i}}}
\def\phie{\phi_{\mathrm{e}}}

\def\Omegar{\Omega_{\mathrm{R}}}
\def\Omegai{\Omega_{\mathrm{I}}}
\def\Omegaar{\Omega_{1\mathrm{R}}}

\def\Omegabr{\Omega_{2\mathrm{R}}}
\def\Omegabi{\Omega_{2\mathrm{I}}}
\def\Omegapr{\Omega_{+\mathrm{R}}}
\def\Omegapi{\Omega_{+\mathrm{I}}}

\allowdisplaybreaks[1]

\begin{document}
\title{Amplifying quantum discord during inflationary magnetogenesis\\
through violation of parity}
\author{Sagarika Tripathy}
\email{E-mail: sagarika@physics.iitm.ac.in}
\affiliation{Centre for Strings, Gravitation and Cosmology,
Department of Physics, Indian Institute of Technology Madras, 
Chennai~600036, India}
\author{Rathul Nath Raveendran}
\email{E-mail: rathulnath.r@gmail.com}
\affiliation{School of Physical Sciences, Indian Association 
for the Cultivation of Science, Kolkata~700032, India}
\author{Krishnamohan Parattu}
\email{E-mail: krishna@iitmandi.ac.in}
\affiliation{Centre for Strings, Gravitation and Cosmology,
Department of Physics, Indian Institute of Technology Madras, 
Chennai~600036, India}
\affiliation{School of Physical Sciences, Indian Institute of Technology 
Mandi, Kamand~175005, India}
\author{L.~Sriramkumar}
\email{E-mail: sriram@physics.iitm.ac.in}
\affiliation{Centre for Strings, Gravitation and Cosmology,
Department of Physics, Indian Institute of Technology Madras,
Chennai~600036, India}

\begin{abstract}
It is well known that, during inflation, the conformal invariance of the 
electromagnetic action has to be broken in order to produce magnetic fields 
of observed strengths today.
Often, to further enhance the strengths of the magnetic fields, parity is 
also assumed to be violated when the fields are being generated. 
In this work, we examine the evolution of the quantum state of the Fourier
modes of the non-conformally coupled and parity violating electromagnetic 
field during inflation.
We utilize tools such as the Wigner ellipse, squeezing parameters and quantum
discord to understand the evolution of the field.
We show that the violation of parity leads to an enhancement of the squeezing 
amplitude and the quantum discord (or, equivalently, in this context, the
entanglement entropy) associated with a pair of opposite wave vectors for one 
of the two states of polarization (and a suppression for the other state of 
polarization), when compared to the case wherein parity is conserved.
We highlight the similarities between the evolution of the Fourier modes of the 
electromagnetic field when parity is violated during inflation and the behavior 
of the modes of a charged, quantum, scalar field in the presence of a constant 
electric field in a de Sitter universe.
We briefly discuss the implications of the results we obtain.
\end{abstract}

\maketitle


\section{Introduction}

Magnetic fields are observed over a wide range of scales in the universe (for 
reviews on magnetic fields, see Refs.~\cite{Grasso:2000wj,Giovannini:2003yn,
Brandenburg:2004jv,Kulsrud:2007an,Subramanian:2009fu,Kandus:2010nw,Widrow:2011hs,
Durrer:2013pga,Subramanian:2015lua,Vachaspati:2020blt}).
They are observed in planets, stars, galaxies, clusters of  galaxies and even 
in the intergalactic medium~(for recent discussions of the various observational 
constraints, see, for example, Refs.~\cite{Paoletti:2019pdi,Vachaspati:2020blt}).
The magnetic fields observed in planets, stars and galaxies can be generated
through astrophysical mechanisms such as batteries (in this context, see, for 
instance, Refs.~\cite{Brandenburg:2004jv,Kulsrud:2007an}).
However, one may need to invoke a cosmological mechanism to explain the
magnetic fields observed in the intergalactic medium~\cite{Neronov:1900zz,
Tavecchio:2010mk,Dolag:2010ni,Dermer:2010mm,Vovk:2011aa,Taylor:2011bn, 
Takahashi:2011ac}.

As is well known, the inflationary paradigm provides a simple and elegant
mechanism for the origin of perturbations in the early universe (see, for 
example, the reviews~\cite{Mukhanov:1990me,Martin:2003bt,Martin:2004um,
Bassett:2005xm,Sriramkumar:2009kg,Baumann:2008bn,Baumann:2009ds,
Sriramkumar:2012mik,Linde:2014nna,Martin:2015dha}).
The scalar and tensor perturbations arise due to quantum fluctuations when 
the Fourier modes are in the sub-Hubble domain during the early stages of
inflation, and they are expected to turn classical as the modes emerge from 
the Hubble radius and evolve onto super-Hubble scales (for discussions in 
this regard, see, for instance, Refs.~\cite{Albrecht:1992kf,Polarski:1995jg,
Kiefer:1998qe,Kiefer:1998jk,Kiefer:2006je,Martin:2007bw,Kiefer:2008ku,
Martin:2012pea,Martin:2012ua,Martin:2015qta}).
The magnetic fields can also be generated in a similar manner.
However, since the standard electromagnetic action is conformally invariant, 
the strengths of the electromagnetic fields produced in such a case will be
rapidly diluted (as $a^{-2}$, with $a$ being the scale factor) during inflation.
Therefore, the conformal invariance of the electromagnetic action has to be 
broken in order to generate magnetic fields of adequate strengths today (see, 
for example, Refs.~\cite{Turner:1987bw,Ratra:1991bn,Bamba:2003av,Bamba:2006ga,
Martin:2007ue,Bamba:2008ja,Demozzi:2009fu,Bamba:2020qdj,Bamba:2021wyx}).
This can be efficiently achieved by coupling the electromagnetic field to 
one or more of the scalar fields that drive inflation~\cite{Bamba:2003av,
Martin:2007ue,Watanabe:2009ct,Kanno:2009ei,Markkanen:2017kmy,Tripathy:2021sfb,
Tripathy:2022iev}.
Interestingly, it has been found that the addition of a parity violating 
term to the action can significantly enhance the strengths of the generated
electromagnetic fields~\cite{Anber:2006xt,Durrer:2010mq,Caprini:2014mja,
Chowdhury:2018mhj,Sharma:2018kgs,Giovannini:2020zjo,Giovannini:2021thf,
Gorbar:2021rlt,Gorbar:2021zlr,Tripathy:2021sfb,Tripathy:2022iev}.

One of the open problems in cosmology today is to understand the 
quantum-to-classical transition of the perturbations generated 
during inflation.
The main challenge in this regard is to identify observable signatures 
that can unequivocally point to the quantum origin of the perturbations.
The evolution of the quantum state associated with the Fourier modes of 
the scalar and tensor perturbations during inflation has been studied 
extensively in the literature (for an intrinsically incomplete list of 
efforts on this topic, see Refs.~\cite{Albrecht:1992kf,Polarski:1995jg,
Kiefer:1998qe,Kiefer:1998jk,Kiefer:2006je,Martin:2007bw,Kiefer:2008ku,
Martin:2012pea,Martin:2012ua,Martin:2015qta,Choudhury:2018ppd,
Brahma:2020zpk,Choudhury:2022mch,Brahma:2023hki}; 
for related discussions in alternative scenarios such as bounces, see 
Refs.~\cite{Battarra:2013cha,Stargen:2016cft,Raveendran:2023dst}).
At the linear order in perturbation theory, these Fourier modes are 
described by time-dependent, quadratic Hamiltonians and, in such 
situations, the unitary evolution operator can be described in terms 
of what are known as the squeezing and rotation 
operators~\cite{Braunstein:2005zz,weedbrook2012gaussian}. 
The evolution of the quantum state of such systems is often tracked 
using the Wigner function, which is a quasi-probability distribution 
in phase space (in this regard, see, for example, 
Refs.~\cite{Hillery:1983ms,case2008wigner}).
The so-called Wigner ellipse is a contour in phase space corresponding
to a given value of the Wigner function.
Usually, the perturbations are assumed to evolve from the ground state, 
in which case, the Wigner ellipse is initially a circle.
As the nomenclature suggests, the squeezing and rotation operators
typically squeeze the Wigner circle into an ellipse and rotate it 
around its center, as the system evolves~\cite{narcowich1990geometry,
Hollowood:2017bil}.

At the linear order in perturbation theory, the Fourier modes associated
with the scalar or tensor perturbations corresponding to the different 
wave numbers evolve independently. 
However, interestingly, one finds that, for a given wave number, say, $k$, 
the Hamiltonian describing the scalar or the tensor perturbations contains
a term that describes an interaction between Fourier modes with the opposite 
wave vectors~${\bmk}$ and~$-{\bmk}$.
As a result, the quantum state associated with these wave vectors proves 
to be entangled~\cite{Lim:2014uea,Martin:2015qta}.
Over the last decade, it has been realized that the notion of quantum discord 
can be utilized as a tool to describe the evolution of the perturbations in 
such situations~\cite{Lim:2014uea,Martin:2015qta,Martin:2021znx}.
Discord is a quintessentially quantum property, i.e. it can be shown to 
be zero for a classical system. 
Moreover, it is more ubiquitous than entanglement, and discordant systems 
contain entangled systems as a subset~\cite{bera2017quantum}.
In other words, a system possessing entanglement will also have a non-zero 
quantum discord, but the converse is not true. 
Since it reflects the quantumness of a system, quantum discord has been 
made use of in cosmology to probe the quantum origin of the cosmological 
perturbations.  
The large quantum discord at the end of inflation has been used to 
argue that cosmological perturbations are indeed placed in a very
quantum state~\cite{Martin:2015qta}.

While the evolution of the quantum state associated with the primordial
scalar and tensor perturbations have been studied in considerable detail, 
we notice that there has only been limited efforts to understand the 
behavior in the case of magnetic fields (in this context, see, for instance, 
Refs.~\cite{Campanelli:2013mea,Maity:2021zng}).
Though there are some similarities between the evolution of scalar or tensor
perturbations and magnetic fields, there can be crucial differences as well. 
In this work, we shall examine the evolution of the quantum state of the
Fourier modes of the non-conformally coupled and parity violating
electromagnetic field during inflation.
Using tools such as the Wigner ellipse, squeezing parameters and quantum 
discord, we shall, in particular, investigate the effects that arise due
to the violation of parity.
Apart from the standard case of slow roll inflation, we shall examine the 
behavior of these measures when there arise departures from slow roll 
inflation.
Specifically, we shall show that the violation of parity amplifies the 
extent of squeezing and quantum discord associated with one of the two 
states of polarization.

This paper is organized as follows.
In the following section, we shall arrive at the action governing the Fourier
modes of the electromagnetic field that is coupled non-conformally to the scalar
field driving inflation.
We shall also consider the effects of an additional term in the action that 
induces the violation of parity.
In Sec.~\ref{sec:qsp}, we shall carry out the quantization of the electromagnetic
modes in the \Sch picture.
We shall also discuss the evolution of the quantum state during inflation.
In Sec.~\ref{sec:ms}, we shall introduce the different measures, such as 
the Wigner ellipse, squeezing parameters and entanglement entropy~(or
quantum discord), that allow us to describe the evolution of the quantum 
state of the electromagnetic field.
In Sec.~\ref{sec:bim}, we shall discuss the behavior of these measures of
the quantum state in specific inflationary models.
In addition to discussing the results in models that permit slow roll 
inflation, we shall discuss the behavior in single and two field models 
that lead to departures from slow roll.
Finally, we shall conclude in Sec.~\ref{sec:s} with a summary of the main
results obtained.
In an appendix, we shall discuss the similarity between the behavior of 
the modes of a parity violating electromagnetic field and a charged scalar 
field in the presence of an electric field in a de Sitter background.
We shall also discuss a few related points in three other appendices.

Before we proceed further, let us clarify a few points regarding the 
conventions and notations that we shall work with. 
We shall work with natural units such that $\hbar=c=1$, and set the reduced 
Planck mass to be $\Mpl=\l(8\,\pi\, G\r)^{-1/2}$.
We shall adopt the signature of the metric to be~$(-,+,+,+)$.
Note that Latin indices will represent the spatial coordinates, except for~$k$ 
which will be reserved for denoting the wave number. 
We shall assume the background to be the spatially flat, 
Friedmann-Lema\^itre-Robertson-Walker~(FLRW) universe described by the following 
line element: 
\begin{equation}
\d s^2=-\d t^2+a^2(t)\,\d {\bm x}^2
=a^2(\eta)\, \l(-\d \eta^2+\d {\bm x}^2\r),\label{eq:FLRW}
\end{equation}
where $t$ and $\eta$ denote cosmic time and conformal time coordinates, 
while $a$ represents the scale factor.
Also, an overdot and an overprime will denote differentiation with respect 
to the cosmic and conformal time coordinates, respectively.
Moreover, $N$ shall represent the number of $e$-folds.
Lastly, $H=\dot{a}/a$ shall represent the Hubble parameter.


\section{Non-conformally coupled electromagnetic fields}\label{sec:nc-emf}

In this section, we shall describe the actions of our interest and express 
them in terms of the Fourier modes of the electromagnetic vector potential.
Later, we shall utilize the reduced action to arrive at the corresponding 
Hamiltonian while discussing the quantization of the electromagnetic modes 
in the \Sch picture.
As we mentioned, we shall be interested in examining a situation wherein the 
electromagnetic field is coupled non-conformally to the scalar field, say, 
$\phi$, that drives inflation.
It proves to be instructive to first discuss non-helical electromagnetic 
fields, before we go on to consider the helical case. 


\subsection{Non-helical electromagnetic fields}

The action describing the non-helical electromagnetic field has the form
\begin{equation} 
S[A^\mu] = -\f{1}{16\,\pi} \int\, \d^4x\, \sqrt{-g}\, J^2(\phi)\, 
F_{\mu\nu}\,F^{\mu\nu}, 
\label{eq:a}
\end{equation}
where $J(\phi)$ denotes the non-conformal coupling function and the field 
tensor~$F_{\mu\nu}$ is expressed in terms of the vector potential $A_\mu$ 
as~$F_{\mu\nu}=( \pa_{\mu}\,A_{\nu}-\pa_{\nu}\,A_{\mu})$.
In a spatially flat, FLRW universe, if we work in the Coulomb gauge wherein 
$A_\eta = 0$ and $\pa_i\,A^i=0$, then the above action reduces to
\begin{equation}
S[A_i] =\f{1}{4\,\pi}\int\d\eta \,\int \d^3 {\bm x}\, J^2(\eta)\,
\l(\f{1}{2}\,A_i'\,A^i{}' -\f{1}{4}\,F_{ij}\,F^{ij}\r),\label{eq:fAi}
\end{equation}
with the spatial indices raised or lowered with the aid of~$\delta^{ij}$
or~$\delta_{ij}$.

Let ${\bmk}$ denote the comoving wave vector and let $\hat{\bmk}$ be the 
corresponding unit vector.
For each vector ${\bmk}$, we can define the right-handed 
orthonormal basis vectors~($\hat{\bm{\varepsilon}}_1^{\bmk},
\hat{\bm{\varepsilon}}_2^{\bmk},
\hat{\bmk}$) which satisfy the relations
\begin{subequations}\label{eq:p-pt}
\begin{eqnarray}
\hat{\bm{\varepsilon}}_1^{\bmk} \cdot \hat{\bm{\varepsilon}}_1^{\bmk}
&= & \hat{\bm{\varepsilon}}_2^{\bmk}\cdot \hat{\bm{\varepsilon}}_2^{\bmk}=1,\;
\hat{\bm{\varepsilon}}_1^{\bmk} \cdot \hat{\bm{\varepsilon}}_2^{\bmk}=0,\\
\hat{\bm{\varepsilon}}_1^{\bmk} \times \hat{\bm{\varepsilon}}_2^{\bmk}&=&\hat{\bmk},\;
\hat{\bmk} \cdot \hat{\bm{\varepsilon}}_1^{\bmk}
= \hat{\bmk} \cdot \hat{\bm{\varepsilon}}_2^{\bmk}=0,\\ 
\hat{\bm{\varepsilon}}_1^{-\bmk} &=&-\hat{\bm{\varepsilon}}_1^{\bmk},\;
\hat{\bm{\varepsilon}}_2^{-\bmk}= \hat{\bm{\varepsilon}}_2^{\bmk}.
\end{eqnarray}
\end{subequations}
Let us denote the components of the polarization vector 
as~$\varepsilon^{\bmk}_{\lambda i}$, where $\lambda=\{1,2\}$ represents 
the two states of polarization of the electromagnetic field.
It can be shown that the components $\varepsilon^{\bmk}_{\lambda i}$ satisfy 
the condition
\begin{equation}
\sum_{\lambda=1}^2 \varepsilon^{\bmk}_{\lambda i}\,
\varepsilon^{\bmk}_{\lambda j} 
=\delta_{ij}-\f{k_i\,k_j}{k^2}   
=\delta_{ij}-\hat{k}_i\,\hat{k}_j,   
\end{equation}
where $\hat{k}_i$ denotes the $i$-th component of unit vector $\hat{\bmk}$.
In terms of the components~$\varepsilon^{\bmk}_{\lambda i}$, we can Fourier 
decompose the vector potential $A_i(\eta,\bm{x})$ in the following manner:
\begin{equation}
A_i(\eta,\bm{x}) = \sqrt{4\,\pi}\, \int  \f{{\d}^3 \bmk}{(2\,\pi)^{3/2}} 
\sum_{\lambda=1}^{2} \varepsilon^{\bmk}_{\lambda i}\, 
\bar{A}^{\lambda}_{\bmk}(\eta)\,\mathrm{e}^{i\,\bmk\cdot\bm{x}}.\label{eq:fAk}
\end{equation}
Since $A_i(\eta,\bm{x})$ and  $\varepsilon^{\bmk}_{\lambda i}$ are real, we 
obtain that 
\begin{equation}
\varepsilon^{\bmk}_{\lambda i}\, \bar{A}^{\lambda}_{\bmk}(\eta)
=\varepsilon^{-\bmk}_{\lambda i}\, 
{\bar{A}^{\lambda\ast}_{-\bmk}}(\eta)  
\end{equation}
and, upon using the properties in~Eqs.~\eqref{eq:p-pt}, this
condition for reality leads to
\begin{equation}
\bar{A}^{1}_{-\bmk}= -\bar{A}^{1\ast}_{\bmk}, \quad 
\bar{A}^{2}_{-\bmk}=\bar{A}^{2\ast}_{\bmk}.\label{eq:rc}
\end{equation}
In terms of the Fourier modes~$\bar{A}_{\bmk}^{\lambda}$, we can 
express the 
action~\eqref{eq:fAi} as follows: 
\begin{eqnarray}
S[\bar{A}_{\bmk}]
= \int \d\eta\, \int \d^3 {\bmk}\,
\sum_{\lambda=1}^{2} J^2(\eta)\,
\l(\f{1}{2}\,\vert\bar{A}_{\bmk}^{\lambda}{}'\vert^2
-\f{k^2}{2}\,\vert\bar{A}_{\bmk}^\lambda\vert^2\r).\nn\\
\label{eq:AAk}
\end{eqnarray}
On varying this action, we obtain the equation of motion governing the modes 
$\bar{A}_{\bmk}^{\lambda}$ to be~\cite{Martin:2007ue,Chowdhury:2018blx}
\begin{equation}
\bar{A}_{\bmk}^{\lambda}{}''
+2\, \f{J'}{J}\,\bar{A}_{\bmk}^{\lambda}{}' 
+k^2\, \bar{A}_{\bmk}^{\lambda} = 0.\label{eq:de-Ak}
\end{equation}


\subsection{Helical electromagnetic fields}

Let us now turn to the case of helical electromagnetic fields.
In general, the action describing helical electromagnetic fields has the
form~\cite{Anber:2006xt,Durrer:2010mq,Caprini:2014mja,Chowdhury:2018mhj,
Sharma:2018kgs,Giovannini:2020zjo,Giovannini:2021thf,Gorbar:2021rlt,
Gorbar:2021zlr,Tripathy:2021sfb,Tripathy:2022iev}
\begin{eqnarray} 
S[A^\mu] &=& -\f{1}{16\, \pi} \int \d^4x\, \sqrt{-g}\, \biggl[J^2(\phi)\,
F_{\mu\nu}\,F^{\mu\nu}\nn\\
& &-\, \f{\gamma}{2}\,I^2(\phi)\,F_{\mu\nu}\widetilde{F}^{\mu\nu}\biggr],
\end{eqnarray}
where $I(\phi)$ represents another coupling function, while $\gamma$ is a 
constant.
The dual field tensor $\widetilde{F}^{\mu\nu}$ is defined as 
$\widetilde{F}^{\mu\nu} = \epsilon^{\mu\nu\alpha\beta}\, F_{\alpha\beta}$, 
with $\epsilon^{\mu\nu\alpha\beta} = (1/\sqrt{-g})\,
\mathcal{A}^{\mu\nu\alpha\beta}$.
The quantity $\epsilon^{\mu\nu\alpha\beta}$ is the totally antisymmetric 
Levi-Civita tensor and $\mathcal{A}^{\mu\nu\alpha\beta}$ is the 
corresponding tensor density with the convention that
$\mathcal{A}^{0123}=1$~\cite{Subramanian:2015lua}.
In a spatially flat, FLRW universe, when working in the Coulomb gauge, the 
above action describing the helical electromagnetic field simplifies to 
\begin{eqnarray}
S[A_i] &=&\f{1}{4\,\pi}\int{\d\eta}\,\int \d^3 {\bm x}\, 
\biggl[J^2(\eta)\,\l(\f{1}{2}\,A_i'\,A^{i}{}' -\f{1}{4}\,F_{ij}\,F^{ij}\r)\nn\\
& &+\,\gamma\, I^2(\eta)\,\epsilon^{ijl}\,A_i'\,\l(\partial_jA_l\r)\biggr].\label{eq:hAi}
\end{eqnarray}
where, as before, the spatial indices are raised or lowered using~$\delta^{ij}$ 
or~$\delta_{ij}$, and $\epsilon^{ijl}$ represents the totally anti-symmetric 
Levi-Civita tensor in three-dimensional Euclidean space.

The Fourier modes of the helical field will be coupled in the basis 
$(\hat{\bm \varepsilon}_1^{\bmk},\hat{\bm{\varepsilon}}_2^{\bmk},\hat{\bmk})$ 
that we had considered in the case of non-helical electromagnetic field. 
To decouple the modes, one can combine the two transverse directions 
$\hat{\bm{\varepsilon}}_1^{\bmk}$ and $\hat{\bm{\varepsilon}}_2^{\bmk}$ to form 
the so-called helicity basis~\cite{Anber:2006xt,Durrer:2010mq,Caprini:2014mja,
Chowdhury:2018mhj,Sharma:2018kgs,Giovannini:2020zjo,Giovannini:2021thf,
Gorbar:2021rlt,Gorbar:2021zlr,Tripathy:2021sfb,Tripathy:2022iev}.
In such a case, we can define an orthonormal basis of vectors 
$(\hat{\bm{\varepsilon}}^{\bmk}_+,\hat{\bm{\varepsilon}}^{\bmk}_-,\hat{\bmk})$, 
where the vectors $\hat{\bm{\varepsilon}}^{\bmk}_\pm$ are defined as 
\begin{equation}
\hat{\bm{\varepsilon}}^{\bmk}_\pm=\f{1}{\sqrt{2}}\,
(\hat{\bm{\varepsilon}}_1^{\bmk}\pm i\,\hat{\bm{\varepsilon}}_2^{\bmk}).
\end{equation}
Using Eqs.~\eqref{eq:p-pt}, one can show that these vectors satisfy the following 
properties:
\begin{subequations}
\begin{eqnarray}
\hat{\bm{\varepsilon}}^{\bmk}_{+} \cdot \hat{\bm{\varepsilon}}^{\bmk\ast}_{+} &=&1,\;
\hat{\bm{\varepsilon}}^{\bmk}_{-} \cdot \hat{\bm{\varepsilon}}^{\bmk\ast}_{-} = 1,\;
\hat{\bm{\varepsilon}}^{\bmk}_{+} \cdot \hat{\bm{\varepsilon}}^{{\bmk}\ast}_{-} = 0,\\
\hat{\bm{\varepsilon}}^{{\bmk}\ast}_\pm &=& \hat{\bm{\varepsilon}}^{\bmk}_\mp,\;
\hat{\bm{\varepsilon}}^{-\bmk}_\pm = -\hat{\bm{\varepsilon}}^{\bmk}_\mp,\;
i\,\hat{\bmk} \times \hat{\bm{\varepsilon}}^{\bmk}_\pm
= \pm \hat{\bm{\varepsilon}}^{\bmk}_\pm.\qquad
\end{eqnarray}
\end{subequations}
Let $\varepsilon^{\bmk}_{\sigma i}$ denote the components of the polarization vector,
with $\sigma=\pm 1$ corresponding to the two helical polarizations in the transverse 
directions of the wave vectors.
It can be established that
\begin{equation}
\sum_{\sigma=\pm} \varepsilon^{\bmk}_{\sigma i}\, 
\varepsilon^{{\bmk}\ast}_{\sigma j} 
=\delta_{ij}-\f{k_i\,k_j}{k^2}   
= \delta_{ij}-\hat{k}_i\,\hat{k}_j.   
\end{equation}
In terms of the components $\varepsilon^{\bmk}_{\sigma i}$ of the
polarization vector, we can decompose the electromagnetic vector 
potential in terms of the Fourier mode functions $\bar{A}^\sigma _{\bmk}$ 
as follows:
\begin{eqnarray}
A_i(\eta,\bm x)
= \sqrt{4\,\pi}\,  \int \f{\d^3{\bmk}}{(2\,\pi)^{3/2}} 
\sum_{\sigma = \pm} \varepsilon^{\bmk}_{\sigma i}\, 
\bar{A}^\sigma _{\bmk}(\eta)\,
\mathrm{e}^{i\,{\bmk}\cdot{\bm x}},\qquad\label{eq:hf-fd}
\end{eqnarray}
Note that $\bar{A}^\sigma_{\bmk}$ can be written in terms 
of $\bar{A}^\lambda_{\bmk}$ as
\begin{equation}
\bar{A}^\sigma _{\bmk}
= \f{1}{\sqrt{2}}\, \l(\bar{A}^1_{\bmk} - i\, \sigma\, \bar{A}^2_{\bmk}\r)
\end{equation}
so that the reality condition~\eqref{eq:rc} becomes
\begin{equation}
\bar{A}^\sigma _{-\bmk}=-\bar{A}^{\sigma\ast}_{\bmk}.\label{eq:rch}
\end{equation}
In terms of the Fourier modes~$\bar{A}^\sigma_{\bmk}$, 
the action~\eqref{eq:hAi} can be expressed as
\begin{eqnarray}\label{eq:a-hc}
S[\bar{A}_{\bmk}^{\sigma}]
&=& \int{\d\eta}\int \d^3{\bmk}\,\,\sum_{\sigma=\pm}\, J^2(\eta)\,
\biggl[\f{1}{2}\,\vert\bar{A}_{\bmk}^{\sigma}{}'\vert^2\nn\\
& &+\,\f{\sigma\, \gamma\, k\, I^2}{2\,J^2}\, \l(\bar{A}_{\bmk}^{\sigma}{}'\,
\bar{A}_{\bmk}^{\sigma\ast}
+\bar{A}_{\bmk}^{\sigma}{}'^{\ast}\,\bar{A}_{\bmk}^{\sigma}\r)
-\f{k^2}{2}\,\vert\bar{A}_{\bmk}^{\sigma}\vert^2\biggr].\nn\\
\end{eqnarray}
On varying this action, we obtain the equation of motion governing the
Fourier modes~$\bar{A}_{\bmk}^\sigma$ to be~\cite{Anber:2006xt,Durrer:2010mq,Caprini:2014mja,
Chowdhury:2018mhj,Sharma:2018kgs,Giovannini:2020zjo,Giovannini:2021thf,
Gorbar:2021rlt,Gorbar:2021zlr,Tripathy:2021sfb,Tripathy:2022iev}
\begin{equation}
\bar{A}_{\bmk}^{{\sigma}''} 
+ 2\,\f{J'}{J}\, \bar{A}_{\bmk}^{\sigma'}
+\bar{\omega}^2\,\bar{A}^\sigma_{\bmk} =0.\label{eq:de-Abark}
\end{equation}
where the quantity $\bar{\omega}^2$ is given by
\begin{equation}
\bar{\omega}^2=k^2+\f{2\,\sigma\,\gamma\,k\,I\,I'}{J^2}.\label{eq:omegabar2}
\end{equation}
We should point out that, in contrast to the non-helical case, the two states
of polarization in the helical case (corresponding to $\sigma=\pm 1$) satisfy 
different equations and hence evolve differently.


\section{Quantization in the \Sch picture}\label{sec:qsp}

In this section, we shall discuss the quantization of the Fourier modes of 
the electromagnetic field in the \Sch picture.


\subsection{Identifying the independent degrees of freedom}\label{sec:i-dof}

To proceed in a manner similar to the analysis of the scalar or tensor 
perturbations described in terms of the associated Mukhanov-Sasaki 
variable, we define the quantity
\begin{equation}
\cA_{\bmk}^{\sigma}=i\,J\,\bar{A}_{\bmk}^{\sigma}.\label{eq:cA-d}    
\end{equation}
We should clarify that the $i$ factor has been introduced so that the reality 
condition~\eqref{eq:rch} becomes
\begin{equation}
\cA^\sigma _{-\bmk}=\cA_{\bmk}^{\sigma}{}^\ast,\label{eq:rcf}    
\end{equation}
mirroring the relation for the Fourier components of the Mukhanov-Sasaki 
variable in the case of the scalar perturbations~\cite{Martin:2015qta}.
In terms of the quantities $\cA_{\bmk}^{\sigma}$, the action~\eqref{eq:a-hc}
can be expressed as
\begin{eqnarray}
S[\cAs] & = & \int{\d\eta}\int \d^3{\bmk}\,\,\sum_{\sigma=\pm}\, 
\biggl[\f{1}{2}\,\vert\cAs'\vert^2 \nn\\ 
& &-\,\f{\kappa}{2}\, \l(\cA_{\bmk}^{\sigma}{}'\, \cA_{\bmk}^{\sigma\ast}
+\cA_{\bmk}^{\sigma}{}'^{\ast}\,\cA_{\bmk}^{\sigma}\r)
- \f{\mu^2}{2}\, \vert\cAs\vert^2\biggr],\qquad\label{eq:Aff}
\end{eqnarray}
where the quantities~$\mu^2$ and~$\kappa$ are given by
\begin{subequations}
\begin{eqnarray}
\mu^2 &=& k^2
-\l(\f{J'}{J}\r)^2+\f{2\,\sigma\,\gamma\, k\,I^2\,J'}{J^3},\\
\kappa &=& \f{J'}{J}-\f{\sigma\, \gamma\, k\,I^2}{J^2}.
\end{eqnarray}
\end{subequations}
The above action has the same structure as the action describing the Fourier 
components of the Mukhanov-Sasaki variable associated with the scalar 
perturbations (in this context, see, for instance, Ref.~\cite{Martin:2015qta}).
But, note that, in the case of the electromagnetic field, the two values 
of~$\sigma$ lead to twice as many degrees of freedom for every~$\bmk$. 
In the non-helical case (i.e. when $\gamma=0$), the above action, in fact, 
reduces {\it exactly}\/ to the form of the action describing the scalar 
perturbations (with the non-conformal coupling function~$J$ replaced by
the scalar pump field~$z$).
However, in the helical case, there arises an important difference with 
the quantities $\mu^2$ and $\kappa$ turning out to be dependent on the 
combination~$(\sigma\,\gamma\,k)$.

With the action describing the Fourier modes of the electromagnetic field at 
hand, we can construct the Hamiltonian associated with each of these modes.
Using the Hamiltonian, we can carry out the quantization of the modes in the 
\Sch picture.
However, note that the reality condition~\eqref{eq:rcf} implies that not all
the Fourier modes~$\cAs$ are independent. 
In order to focus on only the independent degrees of freedom, we divide the 
Fourier space into two parts (such that ${\bmk}$ and $-{\bmk}$ occur in 
different halves) and express the modes in one half in terms of the modes 
in the other half using the relation~\eqref{eq:rcf} (for a similar discussion 
in the case of scalar and tensor perturbations, see, for instance,
Refs.~\cite{Grishchuk:1993,Martin:2012pea,Martin:2015qta,Martin:2019wta,
Agullo:2022ttg,Martin:2021znx}).
The division of the three-dimensional Euclidean space~$\mathbb{R}^{3}$ into two 
can be carried out by any plane passing through the origin.
Therefore, the resultant integral will be over one-half of the Fourier space 
(which we shall denote as~$\mathbb{R}^{3}/2$) so that we have  
\begin{eqnarray}
S[\cAs] & = & \int{\d\eta}\int_{\mathbb{R}^{3}/2} \d^3{\bmk}\,
\sum_{\sigma=\pm}\, \biggl[\vert\cAs'\vert^2 \nn\\ 
& &-\,\kappa\,\l(\cAs\,\cAs'{}^\ast+\cAs^\ast\,\cAs'\r)
- \mu^2\, \vert\cAs\vert^2\biggr].\qquad\label{eq:Af-in}
\end{eqnarray}
Later, we shall be focusing on scenarios wherein $I=J$, with $J'/J$ vanishing at
early times.  
In such situations, due to the term involving $(\sigma\,\gamma\,k)$ in~$\kappa$, 
the above action does not reduce to that of a free harmonic oscillator during the 
initial stages of inflation\footnote{Though, we should hasten to clarify that the 
equation of motion governing~$\cAs$ indeed reduces to that of a free harmonic 
oscillator at such times.}.
We remedy the issue by adding the following total time derivative term to the 
above action:
\begin{align}
-\f{\d}{\d \eta}\l[\f{\sigma\,\gamma\,k\,I^2}{J^2}\,\vert\cAs\vert^2\r].
\label{eq:TD}
\end{align}
In such a case, the resulting action turns out to be 
\begin{eqnarray}
S[\cAs] & = & \int{\d\eta}\int_{\mathbb{R}^{3}/2} \d^3{\bmk}\,
\sum_{\sigma=\pm}\, \biggl[\vert\cAs'\vert^2 \nn\\ 
& &-\,\f{J'}{J}\,\l(\cAs\,\cAs'{}^\ast+\cAs^\ast\,\cAs'\r)
-\bar{\mu}^2\, \vert\cAs\vert^2\biggr],\qquad\label{eq:Af}
\end{eqnarray}
where the quantity $\bar{\mu}^2$ is defined as
\begin{equation}
\bar{\mu}^2=\bar{\omega}^2-\l(\f{J'}{J}\r)^2
\end{equation}
with $\bar{\omega}^2$ given by Eq.~\eqref{eq:omegabar2}.
In App.~\ref{app:cm}, we shall discuss further the reasons for adding the total 
time derivative and working with the modified action.

But, since $\cAs$ is not a real variable, it will not lead to a Hermitian operator 
on quantization. 
Hence, we shall perform the quantization in terms of the real and imaginary parts 
of the variable. 
Let $\cA^{\sigma}_{\bmk\mathrm{R}}/\sqrt{2}$
and~$\cA^{\sigma}_{\bmk\mathrm{I}}/\sqrt{2}$ denote the real and imaginary parts 
of $\cAs$ so that we have
\begin{equation}
\cA_{\bmk}^{\sigma}
=\f{1}{\sqrt{2}}\l(\cA^{\sigma}_{{\bmk\mathrm{R}}}
+i\, \cA^{\sigma}_{\bmk\mathrm{I}}\r).\label{eq:ARI}
\end{equation}
In such a case, we find that the action~\eqref{eq:Af} splits into two 
equivalent terms describing the real and imaginary parts, which implies 
that they evolve independently.
These quantities are governed by the following Lagrangian density 
\textit{in Fourier space}:
\begin{equation}
\mathcal{L}= \f{1}{2}\,\cA'{}^2-\f{J'}{J}\,\cA'\,\cA-\f{\bar{\mu}^2}{2}\,\cA^2.
\label{eq:Lptd-f1}
\end{equation}
where $\cA$ stands for either~$\cA^{\sigma}_{\bmk\mathrm{R}}$
or~$\cA^{\sigma}_{\bmk\mathrm{I}}$. 


\subsection{\Sch equation and the Gaussian ansatz}\label{sec:sv1}

Let us now quantize the system in the \Sch picture.
Given the Lagrangian~\eqref{eq:Lptd-f1}, the canonical conjugate 
momentum~$\cP$ is given by
\begin{equation}
\cP=\cA'-\f{J'}{J}\,\cA.\label{eq:cm}
\end{equation}
The corresponding Hamiltonian density \textit{in Fourier space},\/ can be 
immediately obtained to be\footnote{As with the Lagrangian density 
$\mathcal{L}$ in Fourier space, we shall hereafter refer to~$\mathcal{H}$ 
simply as the Hamiltonian.
Also, the Hamiltonian~$\mathcal{H}$ should not be confused with the 
conformal Hubble parameter, which is often denoted in the same manner.
{\it We do not make use of the conformal Hubble parameter in this paper.}\/}
\begin{equation}
\mathcal{H} = \f{\cP^2}{2}+\f{J'}{J}\,\cP\,\cA
+\f{1}{2}\,\bar{\omega}^2\,\cA^2,
\label{eq:Hamiltonian-AP}
\end{equation}
where $\bar{\omega}^2$ is given by Eq.~\eqref{eq:omegabar2}.
To ensure that, on quantization, the operator corresponding to this
Hamiltonian is Hermitian, as is often done, we shall symmetrize the
classical quantity $(\mathcal{P}\,\mathcal{A})$ and consider the 
corresponding operator to be $(\hat{\mathcal{P}}\,\hat{\mathcal{A}} 
+ \hat{\mathcal{A}}\,\hat{\mathcal{P}})/2$.
In such a case, if $\Psi(\cA,\eta)$ is the wave function describing 
the system, then, on representing the momentum operator~$\hat{\mathcal{P}}$ 
as~$-i\,\pa/(\pa\cA)$, 
the above Hamiltonian leads to the following \Sch 
equation governing the wave function:
\begin{eqnarray}
i\,\f{\pa \Psi}{\pa \eta}
&=&-\f{1}{2}\,\f{\pa^2\Psi}{\pa \cA^2}
-\f{i}{2}\,\f{J'}{J} \l(\Psi+ 2\,\cA\,\f{\pa\Psi}{\pa \cA}\r)
+\f{1}{2}\,\bar{\omega}^2\cA^2\,\Psi.\nn\\
\label{eq:schcA}
\end{eqnarray}
We shall assume that, at very early times, the Fourier modes are in the 
ground state, often referred to as the Bunch-Davies vacuum.
To take into account such an initial condition, we shall assume that the 
wave function is described by the Gaussian ansatz (see, for instance,
Refs.~\cite{Guth:1982ec,cariolaro2015quantum,Martin:2012pea,Katsinis:2023hqn,
Boutivas:2023ksg})
\begin{equation}
\Psi(\mathcal{A},\eta)
=\mathcal{N}(\eta)\, \mathrm{exp}\l[-\Omega(\eta)\,\cA^2/2\r],\label{eq:wfncA}
\end{equation}
where $\mathcal{N}$ and $\Omega$ are, in general, complex quantities.
The normalization of the wave function $\Psi(\cA,\eta)$, viz.
\begin{equation}
\int_{-\infty}^{\infty} {\d} \cA\, \vert\Psi(\cA,\eta)\vert^2 =1,
\end{equation}
immediately leads to the following relation between the 
functions~$\mathcal{N}(\eta)$ and $\Omega(\eta)$:
\begin{equation}
\vert \mathcal{N} \vert=\l(\f{\Omegar}{\pi}\r)^{1/4},\label{eq:N}
\end{equation}
where $\Omegar=(\Omega+\Omega^*)/2$ denotes the real part of the function~$\Omega$.
This implies that $\mathcal{N}$ can be determined (modulo an unimportant 
overall phase factor) if we can obtain~$\Omega$.

Upon substituting the ansatz~\eqref{eq:wfncA} for the wave function~$\Psi(\cA,\eta)$
in the \Sch equation~\eqref{eq:schcA}, we find that the quantity $\Omega$ satisfies 
the differential equation
\begin{equation}
\Omega' = -i\,\Omega^2-2\,\f{J'}{J}\,\Omega+i\,\bar{\omega}^2.\label{eq:ompA}
\end{equation}
In order to solve such a differential equation, let us write
\begin{equation}
\Omega=-i\,\f{g^*}{f^*}, \label{eq:domega}
\end{equation}
where 
\begin{equation}
g= f'-\f{J'}{J}\,f.\label{eq:g} 
\end{equation}
On substituting the above expression for $\Omega$ in Eq.~\eqref{eq:ompA}, 
we arrive at the following equation governing~$f^\ast$:
\begin{equation}
f^{\ast}{}''+\omega^2\,f^*=0, \label{eq:feq}
\end{equation}
where the quantity $\omega^2$ is given by
\begin{equation}
\omega^2 = \mu^2-\kappa'
=\bar{\omega}^2-\f{J''}{J}
=k^2-\f{J''}{J}+\f{2\,\sigma\,\gamma\,k\,I\,I'}{J^2}.\;\;\label{eq:omega2}
\end{equation}
The above differential equation for $f^*$ is identical in form to 
the equation of motion that governs~$\mathcal{A}$ [which can be 
arrived at by substituting the relation~\eqref{eq:cA-d} between~$\cAs$ 
and~$\bar{A}_{\bmk}^{\sigma}$ in Eq.~\eqref{eq:de-Abark}].
In other words, if we know the classical solution for $\cA$ (or, 
equivalently,~$f$), then we can construct the wave 
function~$\Psi(\cA,\eta)$ completely. 


\section{Measures that reflect the evolution of the quantum state}\label{sec:ms}

In this section, we shall discuss the ideas of the Wigner ellipse, squeezing 
parameters and entanglement entropy (or, equivalently, quantum discord), measures 
that help us understand the evolution of the wave function describing the system.


\subsection{Wigner ellipse}\label{sec:we1}

Given a wave function $\Psi(\mathcal{A},\eta)$, the Wigner 
function~$W(\cA,\cP,\eta)$ is 
defined as~\cite{Hillery:1983ms,case2008wigner}
\begin{equation}
W(\cA,\cP,\eta)=\f{1}{\pi}\,
\int_{-\infty}^{\infty} \d y\, \Psi(\cA-y,\eta)\, \Psi^*(\cA+y,\eta)\,
\mathrm{e}^{2\,i\,\cP\,y}.
\end{equation}
For the Gaussian form of $\Psi(\cA,\eta)$ in Eq.~\eqref{eq:wfncA}, we 
can easily obtain the Wigner function to be  
\begin{equation}
W(\mathcal{A},\cP,\eta)
=\f{1}{\pi}\, \mathrm{exp}\l[-\Omegar\,\cA^2-\f{1}{\Omegar}\,(\cP+\Omegai\,\cA)^2\r],
\label{eq:wfn-ga}
\end{equation}
where $\Omegai$ is the imaginary part of $\Omega$, i.e. $\Omegai=(\Omega-\Omega^*)/(2\,i)$.
To visualize the evolution of the Wigner function in the phase 
space $\mathcal{A}$-$\cP$, we can choose to plot the behavior of 
the contour described by the condition
\begin{eqnarray}
\Omegar\,\mathcal{A}^2+\f{1}{\Omegar}\,
\l(\cP+\Omegai\,\mathcal{A}\r)^2=1,\label{eq:phspace}
\end{eqnarray}
which is often referred to as the Wigner ellipse~\cite{case2008wigner,
narcowich1990geometry,Hollowood:2017bil}.

At very early times, if we demand that the wave function $\Psi(\cA,\eta)$
corresponds to the Bunch-Davies vacuum, then the function $f$ is expected
to behave as
\begin{equation}
f\simeq \f{1}{\sqrt{2\,k}}\,\mathrm{e}^{-i\,k\,\eta}.\label{eq:fnh}
\end{equation}
For a power law form of $J$ (say, when $J\propto\eta^{-n}$, where $n$
is a real number), we have, at early times (i.e. as $\eta\to -\infty$)
\begin{equation}
g=f'-\f{J'}{J}\,f 
\simeq -i\,\sqrt{\f{k}{2}}\, \mathrm{e}^{-i\,k\,\eta}.\label{eq:gnh}
\end{equation}
It is useful to note that, for such an initial condition, the Wronskian 
associated with the functions~$f$ and~$g$ is given by
\begin{equation}
\mathcal{W}=f\,g^*-g\,f^*=i.\label{eq:wronskian}
\end{equation}
The above expressions for $f$ and $g$ lead to $\Omegar=k$ and $\Omegai=0$.
If we introduce the following canonical variables which have the same dimension:
\begin{equation}
\bar{\cA}= \sqrt{k}\,\cA,\quad 
\bar{\cP}= \f{\cP}{\sqrt{k}},\label{eq:dcv}
\end{equation}	
then the condition~\eqref{eq:phspace} reduces to 
\begin{equation}
\bar{\cA}^2+\bar{\cP}^2=1.\label{eq:we-bd}
\end{equation}
In other words, at early times, the Wigner ellipse is a circle with 
its centre located at the origin, as in the case of the scalar 
perturbations.


\subsection{Squeezing parameters}\label{sec:sp1}

The squeezing parameters can be related to the components of the 
so-called covariance matrix associated with the wave function.
In terms of the conjugate variables~$\cA$ and~$\cP$, the covariance 
matrix is defined as (see, for example,
Refs.~\cite{cariolaro2015quantum,Martin:2012pea})
\begin{equation}
V = \begin{bmatrix}
k\,\langle \hat{\cA}^2\rangle & \langle
\hat{\cA}\,\hat{\cP}+\hat{\cP}\,\hat{\cA}\rangle/2\\
\langle\hat{\cA}\,\hat{\cP} + \hat{\cP}\,\hat{\cA}\rangle/2 
&\langle \hat{\cP}^2\rangle/k
\end{bmatrix},
\end{equation}
where the expectation values are to be evaluated in the state described by 
the wave function~$\Psi(\cA,\eta)$ [cf. Eq.~\eqref{eq:wfncA}].
The covariance matrix can be expressed in terms of the squeezing amplitude~$\sr$ 
and the squeezing angle~$\varphi$ as follows~\cite{Grishchuk:1993,cariolaro2015quantum,
weedbrook2012gaussian,Martin:2021znx}
\begin{widetext}
\begin{equation}
V = \f{1}{2}\,
\begin{bmatrix}
\cosh\, (2\,\sr) +  \sinh\, (2\,\sr)\,\cos\, (2\, \varphi)  
& \sinh\, (2\,\sr)\,\sin\, (2\, \varphi)\\
\sinh\, (2\,\sr)\,\sin\, (2\, \varphi)   
& \cosh\, (2\,\sr) - \sinh\, (2\,\sr) \,\cos\, (2\, \varphi)
\end{bmatrix}~.\label{eq:cm-sfm}
\end{equation}
\end{widetext}
The shape and orientation of the Wigner ellipse has a one-to-one correspondence 
with the covariance matrix (in this regard, see 
Refs.~\cite{cariolaro2015quantum,weedbrook2012gaussian,Martin:2021znx}; 
in particular, see Ref.~\cite{Raveendran:2022dtb}, App.~A). 
Using the wave function~\eqref{eq:wfncA} and the expressions for $\mathcal{N}$ 
and~$\Omega$ in Eqs.~\eqref{eq:N} and~\eqref{eq:domega}, it can be shown that 
\begin{subequations}\label{eq:f2-g2-in-r}
\begin{eqnarray}
\langle \hat{\cA}^2\rangle
&=&\vert f\vert^2\nn\\ 
&=&\f{1}{2\,k}\,[\cosh\,(2\,\sr)+\sinh(2\,\sr)\,\cos\,(2\,\varphi)],
\label{eq:f2-g2-in-r-1}\nn\\
\\
{\langle \hat{\cP}^2\rangle}
&=&\vert g\vert^2\nn\\
&=&\f{k}{2}\,[\cosh\,(2\,\sr)-\sinh\,(2\,\sr)\,\cos\,(2\,\varphi)],
\label{eq:f2-g2-in-r-2}\nn\\
\\ 
\f{1}{2}\,\langle \hat{\cA}\, \hat{\cP}+ \hat{\cP}\, \hat{\cA}\rangle
&=&\f{1}{2}\,\l(f\,g^{\ast} + f^\ast\,g\r)\nn\\ 
&=&\f{1}{2}\,\sinh\,(2\, \sr)\, \sin\,(2\,\varphi),\label{eq:f2-g2-in-r-3}
\end{eqnarray}
\end{subequations} 
which can be inverted to arrive at
\begin{subequations}\label{eq:rphi-g}
\begin{eqnarray}
\cosh\, (2\,\sr) &=& k\, \vert f\vert^2 +\f{\vert g\vert^2}{k},\label{eq:r-g}\\
\cos\, (2\, \varphi) &=& \f{1}{\sinh\,(2\, \sr)}\,
\l(k\,\vert f\vert^2-\f{\vert g\vert^2}{k}\r),\\
\sin\, (2\, \varphi) &=& \f{1}{\sinh\,(2\, \sr)}\,
\l(f\,g^{\ast} + f^\ast\,g\r).
\end{eqnarray}
\end{subequations}
In other words, if we know the solutions to the classical Fourier modes of 
the electromagnetic field, we can arrive at the squeezing amplitude~$\sr$ 
and squeezing angle~$\varphi$ that describe the evolution of the wave function 
of the quantum system.
We should point out that, since, at early times, $f$ and~$g$ are given by
Eqs.~\eqref{eq:fnh} and~\eqref{eq:gnh}, we have $\cosh\, (2\,\sr)=1$, or, 
equivalently, $\sr=0$.
This essentially indicates that the electromagnetic mode is in its ground 
state at early times.
Note that, in the same limit, the squeezing angle~$\varphi$ is undetermined.


\subsection{Entanglement entropy and quantum discord}\label{sec:qd}

We shall now derive the entanglement entropy and quantum discord that arises 
when we make a particular partition of our system of the non-conformally 
coupled electromagnetic field into two subsystems. 
It can be shown that, when the complete system is in a pure state, quantum 
discord coincides with the entanglement entropy (for a discussion in this 
regard, see, for instance, Refs.~\cite{bera2017quantum,datta2008quantum}). 
Since our system consists only of the electromagnetic field, with the coupling 
to the inflaton being parametrized by time-dependent coefficients, the quantum 
state of the system is in a pure state. 
Therefore, from now on, we shall discuss the entanglement entropy of the system 
and it is to be understood that it is the same as the quantum discord.

In Secs.~\ref{sec:sv1}, \ref{sec:we1} and~\ref{sec:sp1}, we had carried out 
the analysis in terms of the real or imaginary parts of the 
variable~$\cA_{\bmk}^{\sigma}$ defined in Eq.~\eqref{eq:ARI}. 
All these variables are decoupled and hence we could work with a fiducial 
variable representing all of them. 
In terms of these variables, if we start with an initially unentangled state, 
there will be no generation of quantum correlations between the different 
degrees of freedom and, hence, no generation of quantum discord. 
However, we can evaluate the entanglement entropy or quantum discord between the 
${\bm k}$ and $-{\bm k}$ sectors, similar to what has been carried out earlier for 
the scalar perturbations~\cite{Lim:2014uea,Martin:2015qta}. 
In this section, working in the \Sch picture, we shall explicitly derive 
the entanglement entropy of the system that has been partitioned in the same 
manner, i.e. into two sectors of~${\bm k}$ and~$-{\bm k}$.


\subsubsection{Challenges with the modified action}\label{sec:cma}

Recall that we had originally obtained the action~\eqref{eq:Aff} to 
describe the Fourier modes~$\cAs$ of the electromagnetic field.
In order for the action to reduce to that of a free, simple harmonic 
oscillator during the early stages of inflation, we had added the total time 
derivative~\eqref{eq:TD} to eventually arrive at the action~\eqref{eq:Af}.
In this section, we shall point out that there arises a challenge in
working with the action (or, equivalently, the associated conjugate momentum)
to calculate the entanglement entropy between the electromagnetic modes with 
wave vectors~${\bm k}$ and~$-{\bm k}$.

Let us begin by first rewriting the action~\eqref{eq:Af} using the 
relation~\eqref{eq:rcf} between~$\cA^\sigma _{-\bm k}$ and $\cA_{\bmk}^{\sigma}$ 
as follows:
\begin{eqnarray}
S[\cAs,\cAsm]  &=&  \int \d\eta\int_{\mathbb{R}^{3}/2} \d^3{\bmk}\,
\sum_{\sigma=\pm}\,\biggl[\cAs'\,\cAsmp\nn\\ 
& &-\f{J'}{J}\,\l(\cAs'\,\cA^{\sigma}_{-\bm k}+\cAsmp\,\cAs\r)\nn\\
& &-\,\bar{\mu}^2\,\cAs\, \cAsm\biggr].\label{eq:Af-k-k-1}
\end{eqnarray}
The Lagrangian density in Fourier space associated with this action is clearly 
given by 
\begin{eqnarray}
\mathcal{L} &=& \cAs'\,\cAsmp
-\f{J'}{J}\,\l(\cAs'\,\cAsm+ \cAsmp\,\cAs\r)\nn\\
& &-\,\bar{\mu}^2\,\cAs\,\cAsm.
\end{eqnarray}
Therefore, the conjugate momenta, say, $\cP^\sigma_{\bm k}$ 
and~$\cP^\sigma_{-\bm k}$, associated with the variables~$\cAs$ 
and~$\cAsm$ are given by 
\begin{subequations}\label{eq:pdef}
\begin{eqnarray}
\cP^\sigma_{\bm k} &=& \f{\pa \mathcal{L}}{\pa \cAsmp}=\cAs'-\f{J'}{J}\, \cAs,\\
\cP^\sigma_{-\bm k} &=& \f{\pa \mathcal{L}}{\partial \cAs'}
=\cAsmp-\f{J'}{J}\, \cAsm.
\end{eqnarray}
\end{subequations}
These conjugate momenta satisfy the relation
\begin{equation}
\cP^\sigma_{-\bm k}=\cP_{\bm k}^{\sigma}{}^\ast, \label{eq:pcf}
\end{equation}
which is akin to the relation~\eqref{eq:rcf} between~$\cAs$ and $\cAsm$.
The Hamiltonian of the system containing the two subsystems $\bm {k}$ and $\bm {-k}$ 
can be obtained to be (for a similar discussion in the case of scalar perturbations,
see, for instance, Ref.~\cite{Martin:2015qta})
\begin{eqnarray}
\mathcal{H}&=& \cP_{\bm k}^\sigma\, \cP_{-\bm k}^\sigma
+\f{J'}{J}\,\l(\cAs\, \cP_{-\bm k}^\sigma+ \cAsm\, \cP_{\bm k}^\sigma\r)
+\bar{\omega}^{2}\cAs\,\cAsm,\nn\\\label{eq:Hkmk}
\end{eqnarray}
where the quantity $\bar{\omega}^2$ is given by Eq.~\eqref{eq:omegabar2}.

The conjugate variables~$(\cAs, \cP^\sigma_{\bm k})$ 
and~$(\cAsm, \cP^\sigma_{-\bm k})$ that appear in the above 
Hamiltonian are {\it not}\/ real.
Hence, they will not turn out to be Hermitian when they are elevated to be 
operators on quantization.
Motivated by the approach that has been adopted in the case of the scalar
perturbations (in this context, see Refs.~\cite{Martin:2015qta,Martin:2019wta}), 
we can define the new quantities~$(x^\sigma_{\bm k},p_{\bm k}^{\sigma})$ 
in terms of $(\cAs, \cP^{\sigma}_{\bm k})$ 
and $(\cAsm, \cP^{\sigma}_{-\bm k})$ as follows:
\begin{subequations}\label{eq:xp-Ap}
\begin{eqnarray}
x^\sigma_{\bm k} &=& \f{1}{2}\, \l(\cAs + \cAsm\r)
+\f{i}{2\,\bar{\omega}}\, \l(\cP^{\sigma}_{\bmk}-\cP^{\sigma}_{-\bmk}\r),\label{eq:x_Ap}\\ 
p^\sigma_{\bm k} &=& \f{1}{2}\, \l(\cP^{\sigma}_{\bm k} + \cP^{\sigma}_{-\bm k}\r)
-\f{i\,\bar{\omega}}{2}\, \l(\cAs-\cAsm\r),\label{eq:p-Ap}
\end{eqnarray}
\end{subequations}
and quantize the system in terms of these new variables.
Note that, in the non-helical case wherein $\gamma=0$, the quantity~$\bar{\omega}^2$ 
[cf. Eq.~\eqref{eq:omegabar2}] reduces to~$k^2$, and hence the new variables are similar
to those encountered in the scalar case (with the non-conformal coupling function~$J$ 
replaced by the pump field~$z$).
However, we find that, in the helical case, i.e. when $\gamma$ is non-zero, the 
quantity $\bar{\omega}^2$ may not remain positive definite over some domains in time,
and hence the quantity $\bar{\omega}$ may turn out to be imaginary.
This implies that the quantities $(x^\sigma_{\bm k},p_{\bm k}^{\sigma})$ will 
{\it not}\/ remain real, and hence cannot be utilized for carrying out the 
quantization of the system\footnote{One simple way to overcome this difficulty 
would be to replace~${\bar \omega}$ in Eqs.~\eqref{eq:xp-Ap} by either $k$ or 
$\vert {\bar \omega}\vert$.
But, the resulting Hamiltonians turn out to be rather cumbersome to deal with. 
It would be worthwhile to examine if one can construct other canonical variables 
which remain real and can be utilized to quantize the system.}.
Therefore, in what follows, we shall turn to the original action~\eqref{eq:Aff} 
for quantization and the evaluation of the entanglement entropy between the
electromagnetic modes with wave vectors $\bmk$ and $-\bmk$.


\subsubsection{Working with the original action}

Note that the original action~\eqref{eq:Af-in} can be expressed as
\begin{eqnarray}
S[\cAs,\cAsm]  &=&  \int \d\eta\int_{\mathbb{R}^{3}/2} \d^3{\bmk}\,
\sum_{\sigma=\pm}\,\biggl[\cAs'\,\cAsmp\nn\\ 
& &-\kappa\,\l(\cAs'\,\cA^{\sigma}_{-\bm k}+\cAsmp\,\cAs\r)
-\mu^2\,\cAs\, \cAsm\biggr]\label{eq:Af-k-k}\nn\\
\end{eqnarray}
so that the associated Lagrangian density in Fourier space is given by
\begin{eqnarray}
\mathcal{L} &=& \cAs'\,\cAsmp
-\kappa\,\l(\cAs'\,\cAsm+ \cAsmp\,\cAs\r)\nn\\
& &-\,\mu^2\,\cAs\,\cAsm.
\end{eqnarray}
The conjugate momenta associated with the variables~$\cAs$ and~$\cAsm$
can be easily obtained to be
\begin{subequations}\label{eq:pdef1}
\begin{eqnarray}
\cP^\sigma_{\bm k} &=& \f{\pa \mathcal{L}}{\pa \cAsmp}
={\cAs'}-\kappa\, \cAs,\\
\cP^\sigma_{-\bm k} &=& \f{\pa \mathcal{L}}{\pa \cAs'}
=\cAsmp-\kappa\, \cAsm.
\end{eqnarray}
\end{subequations}
which correspond to the conjugate momentum in Eq.~\eqref{eq:cm1}.
In such a case, we find that the Hamiltonian of the system can be expressed 
as
\begin{eqnarray}
\mathcal{H}&=& \cP_{\bm k}^\sigma\, \cP_{-\bm k}^\sigma
+ \kappa\,\l(\cAs\, \cP_{-\bm k}^\sigma+ \cAsm\, \cP_{\bm k}^\sigma\r)
+\widetilde{\omega}^{2}\,\cAs\,\cAsm,\nn\\
\label{eq:Hkmk2}
\end{eqnarray}
where the quantity $\widetilde{\omega}^2$ is given by
\begin{equation}
\widetilde{\omega}^2=k^2\,\l({1+\f{\gamma^2\,I^4}{J^4}}\r).\label{eq:omegatilde2}
\end{equation}
We should point out here that, in contrast to the quantity~${\bar \omega}^2$ 
[cf. Eq~\eqref{eq:omegabar2}] which we had encountered in the 
Hamiltonian~\eqref{eq:Hkmk} earlier, the quantity~$\widetilde{\omega}^2$ 
that appears in the above Hamiltonian is clearly positive definite. 

We can now make use of the transformations~\eqref{eq:xp-Ap} with~${\bar \omega}$ 
replaced by~$\widetilde{\omega}$ to arrive at the new set of real 
variables~$(x^\sigma_{\bm k},p_{\bm k}^{\sigma})$.
In terms of these variables, the Hamiltonian~\eqref{eq:Hkmk2} of the system turns 
out to be 
\begin{eqnarray}
{\mathcal H} 
&=& \f{1}{2}\,\l(p_{\bm k}^\sigma\, p_{\bm k}^\sigma 
+ p_{-\bmk}^\sigma\, p_{-\bmk}^\sigma\r)
+\kappa\,\l(x_{\bmk}^\sigma\, p_{-\bmk}^\sigma+x_{-\bmk}^\sigma\, p_{\bmk}^\sigma\r)\nn\\
& & +\,\f{\widetilde{\omega}^{2}}{2}\, 
\l(x_{\bmk}^\sigma\, x_{\bmk}^\sigma+x_{-\bmk}^\sigma\, x_{-\bmk}^\sigma\r).
\end{eqnarray}
For convenience, we shall hereafter refer to $x_{\bmk}^\sigma$ and $x_{-\bmk}^\sigma$ 
simply as~$x_1$ and~$x_2$.
The \Sch equation describing the system can then be written as
\begin{eqnarray}
i\,\f{\pa \Psi}{\pa \eta} 
&=& -\f{1}{2}\,\l(\f{\pa^2\Psi}{\pa x_1^2}+\f{\pa^2\Psi}{\pa x_2^2}\r)
-i\, \kappa\,\l(x_1\, \f{\pa \Psi}{\pa x_2}+x_2\, \f{\pa \Psi}{\pa x_1}\r)\nn\\ 
& &+ \f{\widetilde{\omega}^{2}}{2}\, (x_1^2+x_2^2)\,\Psi. \label{eq:Sch-eqn}
\end{eqnarray}
As we had done earlier [cf. Eq.~\eqref{eq:wfncA}], we can consider the following 
Gaussian ansatz for the wave function describing the system:
\begin{eqnarray}
\Psi(x_1,x_2,\eta)\ &= & \mathcal{N}(\eta)\,
\mathrm{exp}\Biggl[-\f{1}{2}\,\Omega_1(\eta)\, \l(x_1^2+x_2^2\r)\nn\\
& & -\,\Omega_2(\eta)\,x_1\,x_2\biggr],\label{eq:ga2}  
\end{eqnarray}
where, evidently, $\mathcal{N}$ is a {\it new}\/, suitable, normalization factor.
The normalization of the wave function leads to the condition
\begin{equation}
\vert \mathcal{N}\vert 
= \l(\f{\Omegaar^2-\Omegabr^2}{\pi^2}\r)^{1/4},\label{eq:cN2}
\end{equation}
where $\Omegaar=(\Omega_1+\Omega_1^\ast)/2$ and $\Omegabr=(\Omega_2+\Omega_2^\ast)/2$ 
are the real parts of the quantities~$\Omega_1$ and~$\Omega_2$.

On substituting the wave function~\eqref{eq:ga2} in the \Sch 
equation~\eqref{eq:Sch-eqn}, we find that the quantities
$\Omega_1$ and $\Omega_2$ satisfy the following differential
equations:
\begin{subequations}
\begin{eqnarray}
\Omega_1' &=& -i\,\l(\Omega_1^2+\Omega_2^2\r)- 2\,\kappa\, 
\Omega_2 +i\,\widetilde{\omega}^2,\label{eq:de-omega-kk}\\
\Omega_2' &=& -2\,i\, \Omega_1\, \Omega_2- 2\, \kappa\, 
\Omega_1.\label{eq:de-omega-k-k}  
\end{eqnarray}
\end{subequations}
If we now define~$\Omega_+=\Omega_1+\Omega_2$, upon combining the 
above equations for~$\Omega_1$ and~$\Omega_2$, it is easy to show 
that the quantity $\Omega_+$ satisfies the equation
\begin{eqnarray}
\Omega_+' &=& -i\,\Omega_+^2-2\,\kappa\,\Omega_++i\, \widetilde{\omega}^2,
\end{eqnarray}
where, recall that, the quantity $\widetilde{\omega}^2$ is given by 
Eq.~\eqref{eq:omegatilde2}.
Hereafter, we shall restrict ourselves to the situations wherein $I=J$, in 
which case $\widetilde{\omega}^2=k^2\,(1+\gamma^2)$, i.e. it reduces to a 
constant.
In such situations, it is also straightforward to establish that, if we 
define $\Omega_{-} =\Omega_1-\Omega_2$, then the above equations for~$\Omega_1$ 
and~$\Omega_2$ imply that $\Omega_-=\widetilde{\omega}^2/\Omega_+$.
Note that the above equation satisfied by~$\Omega_+$ is the same as 
Eq.~\eqref{eq:omega1} that governs~$\Omega$. 
Therefore, if we use the definition~\eqref{eq:domega} for~$\Omega_+$,
with~$g$ given by Eq.~\eqref{eq:g1}, then $f^\ast$ satisfies the equation 
of motion~\eqref{eq:feq}.
In other words, as with the wave function $\Psi(\cA,\eta)$ 
[cf. Eq.~\eqref{eq:wfncA}] that describes the unentangled state associated 
with the wave number~$k$, the wave function $\Psi(x_1,x_2,\eta)$ 
[cf. Eq.~\eqref{eq:ga2}] that carries information about the interaction 
between the wave vectors $\bmk$ and $-\bmk$ can also be completely expressed 
in terms of the classical solutions to the Fourier modes of the electromagnetic 
field.
With $\Omega_+$ and $\Omega_-$ at hand, we can obtain $\Omega_1$ and
$\Omega_2$ using the relations
\begin{subequations}\label{eq:om1-om2-omega}
\begin{eqnarray}
\Omega_1 &=&\f{1}{2}\,(\Omega_++\Omega_-)
=\f{1}{2\,\Omega_+}\,(\Omega_+^2+\widetilde{\omega}^2),
\label{eq:om1-om}\\
\Omega_2 &=&\f{1}{2}\,(\Omega_+-\Omega_-) 
=\f{1}{2\,\Omega_+}\,(\Omega_+^2-\widetilde{\omega}^2),\label{eq:om2-om} 
\end{eqnarray}
\end{subequations}
which, in turn, allow us to construct the wave function~$\Psi(x_1,x_2,\eta)$.


\subsubsection{Derivation of the entanglement entropy}\label{subsec:d-ee}

Note that the {\it complete}\/ wave function~$\Psi(x_1,x_2,\eta)$ of the 
system of our interest describes a pure state and hence does not possess 
any entanglement entropy.
We shall trace one of the two degrees of freedom to arrive at the reduced 
density matrix and evaluate the corresponding entanglement entropy\footnote{As
is well known, the entanglement entropy of a bipartite system proves to be the 
same, independent of which of the two parts of the system is traced over.}.
The reduced density matrix, obtained by tracing out the degrees of freedom
associated with the variable~$x_1$, is defined as
\begin{equation}
\rho_{\mathrm{red}}(x_2,x_2',\eta)
=\int_{-\infty}^{\infty}\d x_1\, \Psi(x_1,x_2,\eta)\,\Psi^\ast(x_1,x_2',\eta),
\end{equation}
with the wave function $\Psi(x_1,x_2,\eta)$ given by Eq.~\eqref{eq:ga2}.
The Gaussian integral over $x_1$ can be easily evaluated to arrive at the 
reduced density matrix
\begin{eqnarray}
\rho_{\mathrm{red}}(x_2,x_2',\eta)
&=& \l\vert\mathcal{N}\r\vert^2\,\sqrt{\f{\pi}{\Omegaar}}\,
\mathrm{exp}\biggl[-\f{\alpha}{2}\,\l(x_2^2+x_2'{}^{2}\r)\nn\\
& &+\,\beta\, x_2\, x_2'\biggr],\label{eq:rdm}
\end{eqnarray}
where~$\vert\mathcal{N}\vert$ is given by Eq.~\eqref{eq:cN2}, while~$\alpha$ 
and~$\beta$ are {\it real}\/ quantities which are given by the expressions
\begin{subequations}\label{eq:alpha-beta}
\begin{eqnarray}
\alpha &=& \Omega_1-\f{\Omega_2^2}{2\,\Omegaar}\nn\\
&=&\f{1}{2\,\Omegaar}\,\l[2\,\Omegaar^2-\l(\Omegabr^2-\Omegabi^2\r)\r],\\
\beta &=& \f{\vert\Omega_2\vert^2}{2\,\Omegaar},
\end{eqnarray}
\end{subequations}
with $\Omegabi=(\Omega_2-\Omega_2^*)/(2\,i)$ denoting the imaginary part 
of~$\Omega_2$.
It is also useful to note here that $(\alpha^2-\beta^2)=\widetilde{\omega}^2$.

Our aim is to now calculate the entanglement entropy associated with the 
above reduced density matrix.
Since the system of our interest behaves as a time-dependent oscillator, the 
entanglement entropy of the system, say, $\mathcal{S}$, can be expressed as
\begin{equation}
\mathcal{S} =- \sum_{n=0}^\infty p_n\, \ln\, p_n,\label{eq:ee}
\end{equation}
where $p_n$ denotes the probability of finding the system in the $n$-th energy
eigen state of the oscillator.
Since the entanglement entropy is the same as the quantum discord for a pure
state, we shall hereafter refer to $\mathcal{S}$ above as quantum discord~$\delta$, 
in the manner it is often done in the context of the scalar 
perturbations~\cite{Martin:2015qta,Martin:2019wta}. 
The eigen values $p_n$ of the reduced density matrix 
$\rho_{\mathrm{red}}(x_2,x_2',\eta)$ are determined by the relation (for an early
discussion, see Ref.~\cite{Srednicki:1993im}; for a recent discussion in this 
context, see, for instance, Ref.~\cite{Katsinis:2023hqn,Boutivas:2023ksg})
\begin{equation}
\int_{-\infty}^{\infty} \d x_2'\, \rho_{\mathrm{red}}(x_2,x_2',\eta)\, 
\psi_n(x_2',\eta) =p_n\,\psi_n(x_2).\label{eq:In}
\end{equation}
The quantities $\psi_n(x)$ are the energy eigen states of the harmonic oscillator
with unit mass and frequency $\widetilde{\omega}$, and are given by
\begin{eqnarray}
\psi_n(x) = \f{1}{2^n\,n!}\,\l(\f{\widetilde{\omega}}{\pi}\r)^{1/4}\,
H_n\l(\sqrt{\widetilde{\omega}}\,x\r)\;
{\mathrm{e}}^{-\widetilde{\omega}\,x^2/2},
\label{eq:sho-ees}
\end{eqnarray}
where the function $H_n(z)$ denotes the Hermite polynomial.
With the density matrix $\rho_{\mathrm{red}}(x_2,x_2',\eta)$ and the wave 
function $\psi_n(x)$ at hand [as given by Eqs.~\eqref{eq:rdm} and~\eqref{eq:sho-ees}],
it is straightforward to carry out the integral~\eqref{eq:In} and determine
the probability $p_n$ to be~\cite{gradshteyn2007table}
\begin{equation}
p_n=(1-\xi)\,\xi^n,\label{eq:pn}
\end{equation}
where $\xi$ is given by
\begin{equation}
\xi=\f{\beta}{\widetilde{\omega}+\alpha}.\label{eq:xi}
\end{equation}
With the help of the above expression for $p_n$, we can carry out the sum 
in the definition~\eqref{eq:ee} of the entanglement entropy (or quantum discord) to 
arrive at the following result in terms of $\xi$ (in this context, see, for
example, Refs.~\cite{Srednicki:1993im,Katsinis:2023hqn,Boutivas:2023ksg}):
\begin{equation}
\delta =-\ln\, (1-\xi) -\f{\xi}{1-\xi}\, \ln \xi.\label{eq:ee-f}
\end{equation}
An equivalent expression that is more convenient for later numerical evaluation 
in specific inflationary models is given by {(for 
further details, see App.~\ref{app:d-ee})}
\begin{equation}
\delta= \l(1+\f{y}{2}\r)\,\ln\l(1+\f{2}{y}\r) +\ln\l(\f{y}{2}\r),
\label{eq:ee-ff}
\end{equation}
where $y$ is related to $\xi$ as follows:
\begin{equation}
y=\f{2\,\xi}{1-\xi}.
\label{eq:def-y}
\end{equation}
Upon using Eqs.~\eqref{eq:om1-om2-omega}, \eqref{eq:alpha-beta}, 
\eqref{eq:xi} and~\eqref{eq:def-y}, we find that the quantity $y$ 
can be expressed as 
\begin{equation}
y  = \f{\l(\widetilde{\omega} - \Omegapr\r)^2 
+ \Omegapi^2}{2\, \widetilde{\omega}\, \Omegapr},\label{eq:y1}
\end{equation}
where $\Omegapr=(\Omega_++\Omega_+^\ast)/2$ and 
$\Omegapi=(\Omega_+-\Omega_+^\ast)/(2\,i)$ denote the real and 
imaginary parts of~$\Omega_+$.
If we further use Eq.~\eqref{eq:domega}, we obtain that
\begin{equation}
y= \f{\l(1-2\,\widetilde{\omega}\,\vert f \vert^2\r)^2
+\l(f\,g^\ast+g\,f^\ast\r)^2}{4\,\widetilde{\omega}\,\vert f \vert^2},\label{eq:y}
\end{equation}
with $\widetilde{\omega}^2$ given by Eq.~\eqref{eq:omegatilde2} and 
$g$ being defined as in Eq.~\eqref{eq:g1}.
Note that the quantities in the above expression for~$y$ depend on the 
non-conformal coupling function~$J$ (recall that we have set $I=J$)
and the solution~$f$. 
In other words, we can evaluate the quantum discord~$\delta$ if we know 
the classical solutions to the Fourier modes of the electromagnetic vector
potential.
In the following section, we shall use the expressions~\eqref{eq:ee-ff} 
and~\eqref{eq:y} to evaluate the quantum discord in different models of 
inflation.
We should mention that, in App.~\ref{app:ad-qd}, we have provided an 
alternative derivation of the quantum discord, obtained from the 
covariance matrix of the system.
Moreover, we ought to clarify that, since we are focusing on a {\it single}\/
wave number of the electromagnetic field, we do not encounter any divergences
when we calculate the entanglement entropy or quantum discord.
As is well known, the divergences are encountered only when we sum (or, integrate)
over all the wave numbers associated with the field.

However, before we proceed to calculate the evolution of the different  
measures describing the state of the electromagnetic field in specific 
inflationary scenarios, we ought to make a few clarifying remarks.
Earlier, when we had focused on a single wave number of the electromagnetic
field (in Secs.~\ref{sec:sv1}, \ref{sec:we1} and~\ref{sec:sp1}), we had 
worked with the modified action~\eqref{eq:Af}, which corresponds to working
with the conjugate momenta~\eqref{eq:pdef} or~\eqref{eq:cm}.
In contrast, when calculating the quantum discord between the
electromagnetic modes with the wave vectors $\bmk$ and $-\bmk$, we 
have instead worked with the original action~\eqref{eq:Aff}, which
leads to the conjugate momenta~\eqref{eq:pdef1} or~\eqref{eq:cm1}.
We have already described the reason for doing so, viz. the fact that 
the transformations~\eqref{eq:xp-Ap} do not lead to real variables when
$\bar{\omega}^2$ [given by Eq.~\eqref{eq:omegabar2}] proves to be negative.
We should also caution that, when $\bar{\omega}^2$ turns negative, the 
derivation of the entanglement entropy we have outlined above---which 
is based on the wave function $\psi_n(x)$ describing the normal oscillator 
[cf. Eq.~\eqref{eq:sho-ees}]---may not apply.

There is yet another point that we need to make at this stage of our discussion.
In the non-helical case, $\bar{\omega}$ reduces to the wave number~$k$ and hence
the above-mentioned problems do not arise.
Also, in such a situation, the actions~\eqref{eq:Af-k-k-1} and~\eqref{eq:Af-k-k}
[and, hence, the corresponding conjugate momenta~\eqref{eq:pdef} and~\eqref{eq:pdef1}]
reduce to the same form and, in fact, exactly resemble the action describing the 
scalar perturbations, as we have already mentioned. 
Under this condition, on using the expressions~\eqref{eq:f2-g2-in-r} 
from the previous section, we find that the quantity~$y$ as defined 
in Eq.~\eqref{eq:y} can be written in terms of the squeezing 
amplitude~$\sr$ as follows:
\begin{equation}
y=\cosh\,(2\, \sr) -1.\label{eq:yr}
\end{equation}
On substituting this relation in Eq.~\eqref{eq:ee-ff}, we can readily obtain an 
expression for quantum discord~$\delta$ in terms of the squeezing amplitude~$\sr$
in the non-helical case.
In fact, at late times during inflation, since the squeezing amplitude~$\sr$ proves
to be large, we have $y\propto \exp\,(2\,\sr)$ so that the quantum discord behaves 
as $\delta \propto 2\,\sr$ [cf. Eqs.~\eqref{eq:yr} and~\eqref{eq:ee-ff}], as in the 
case of the scalar perturbations~\cite{Martin:2015qta}.
However, we should clarify here that, for the helical fields, we do not have 
an explicit expression that relates the quantum discord~$\delta$ and the 
squeezing amplitude~$\sr$.
Therefore, we shall work with Eqs.~\eqref{eq:ee-ff} and~\eqref{eq:y} to evaluate
quantum discord for the parity violating electromagnetic fields.
In the next section, when we discuss the numerical results in specific inflationary 
models, we shall see that, even in the helical case, the quantum discord 
has a similar relation to the squeezing amplitude (i.e. $\delta\propto 2\,\sr$)
at late times.


\section{Behavior in different inflationary scenarios}\label{sec:bim}

With various tools to describe the evolution of the quantum state of the 
electromagnetic modes at hand, let us examine the evolution of the state 
in some specific situations.
In the following sections, we shall examine the evolution of the quantum
state in simple situations involving slow roll as well as in non-trivial
scenarios permitting some departures from slow roll.
We shall assume that~$I=J$ and focus on the helical case. 
Evidently, the results for the non-helical case can be obtained by 
considering the limit wherein $\gamma$ vanishes.


\subsection{In de Sitter inflation}\label{sec:dsi}

Let us first discuss the often considered de Sitter case as it permits
analytical solutions.
Evidently, we shall require a form of $J(\eta)$ in order to make progress.
The non-conformal coupling function that breaks the conformal invariance 
of the standard electromagnetic action is typically assumed to be of the 
following form~\cite{Subramanian:2009fu,Subramanian:2015lua,Sharma:2018kgs,
Tripathy:2021sfb,Tripathy:2022iev}:
\begin{equation}
J(\eta)= \l[\f{a(\eta)}{a(\ee)}\r]^n, \label{eq:J}
\end{equation}
where $\ee$ is the conformal time at the end of inflation and the 
parameter~$n$ is a real number.
Note that the non-conformal coupling function reduces to unity at the end 
of inflation.
As is well known, in the de Sitter case, the above coupling function leads 
to a scale invariant spectrum of the magnetic field for~$n=2$ and~$n=-3$.
We shall restrict our discussion to~$n = 2$ throughout this work in order 
to avoid the issue of backreaction~(in this context, see, for instance, 
Refs.~\cite{Markkanen:2017kmy,Tripathy:2021sfb}).

Recall that, in de Sitter inflation, the scale factor describing the FLRW
universe is given by $a(\eta)=-1/(\HI\,\eta)$, where $\HI$ is the Hubble 
parameter which is a constant.
In such a case, $J$ is given  by
\begin{equation}
J(\eta) = \l(\f{\eta}{\ee}\r)^{-n}\label{eq:J-dS}
\end{equation}
so that 
\begin{equation}
\f{J'}{J}=-\f{n}{\eta},\quad \f{J''}{J}= \f{n\,(n+1)}{\eta^2}
\end{equation}
and, hence, the function $f$ which describes the wave function $\Psi(\cA,\eta)$
[cf. Eqs.~\eqref{eq:wfncA}, \eqref{eq:domega} and~\eqref{eq:feq}] satisfies 
the differential equation
\begin{equation}
f''+ \l[k^2 - \f{2\,\sigma\,\gamma\,k\,n}{\eta}
- \f{n\,(n+1)}{\eta^2}\r]\,f = 0.\label{eq:cA-h-de-sp}
\end{equation}
The solution to this differential equation which satisfies the Bunch-Davies 
initial conditions at early times can be written as follows (for recent 
discussions, see, for example, Refs.~\cite{Sharma:2018kgs,Tripathy:2021sfb}):
\begin{equation}
f(\eta) =\f{1}{\sqrt{2\,k}}\, \mathrm{e}^{-\sigma\,\pi\,n\,\gamma/2}\,
W_{i\,\sigma\,n\,\gamma,\nu}(2\,i\,k\,\eta),\label{eq:cA-h}
\end{equation}
where $\nu=n+(1/2)$ and $W_{\lambda,\nu}(z)$ denotes the Whittaker 
function~\cite{gradshteyn2007table}.
We find that, as $(-k\,\eta) \to \infty$, the above function~$f$ and the 
quantity $g=f'-(J'/J)\,f$ reduce to the asymptotic forms in Eqs.~\eqref{eq:fnh} 
and~\eqref{eq:gnh}, as required.
We should mention that, for a range of values of the Hubble parameter $\HI$, 
the parameter~$\gamma$ and $n\simeq 2$, the resulting spectrum of the magnetic 
field proves to be nearly scale invariant and consistent with the current 
constraints from observations~\cite{Tripathy:2021sfb,Tripathy:2022iev}.

We can now arrive at the squeezing amplitude~$\sr$ upon using the 
solution~\eqref{eq:cA-h} for the mode function~$f$ in the 
expression~\eqref{eq:r-g}.
It is useful to note that, in the domain $z\ll 1$, the Whittaker 
function $W_{\lambda,\mu}(z)$ behaves as~\cite{Gradshteyn:1702455,
Mathematica}
\begin{eqnarray}
W_{\lambda,\mu}(z)
& \simeq & \f{\Gamma(-2\,\mu)}{\Gamma(\tfrac{1}{2} -\lambda -\mu)}\, 
z^{(1/2)+\mu}\nn\\
& &+\, \f{\Gamma(2\,\mu)}{\Gamma(\tfrac{1}{2} - \lambda +\mu)}\,
z^{(1/2)-\mu}.\label{eq:Wfn-s}
\end{eqnarray}
Also, the derivative of the Whittaker function can be expressed 
as~\cite{Gradshteyn:1702455,Mathematica}
\begin{equation}
\f{\d W_{\lambda,\mu}(z)}{\d z} = \l(\f{1}{2}-\f{\lambda}{z}\r)\,
W_{\lambda,\mu}(z)-\f{1}{z}\,W_{1+\lambda,\mu}(z)
\end{equation}
and we can use the following recursion relation to further simplify 
the expression:
\begin{equation}
W_{\lambda,\mu}(z)
=\sqrt{z}\,W_{\lambda-\tfrac{1}{2},\mu-\tfrac{1}{2}}(z)
+\l(\f{1}{2}-\lambda+\mu\r)\,W_{\lambda-1,\mu}(z).
\end{equation}
When $n=2$, it is easy to show that, at late times (i.e. as $\eta
\to 0$)
\begin{subequations}
\begin{eqnarray}
k\,\vert f\vert^2 
&\simeq& \f{9\, (1-\mathrm{e}^{-4\,\pi\,\sigma\,\gamma})}{8\,\pi\,
\sigma\,\gamma\,(1+ 5\,\gamma^2+4\,\gamma^4)}\,\l(\f{1}{-k\,\eta}\r)^4,\;\;\\
\f{\vert g\vert^2}{k} 
&\simeq& \f{9\, (1-\mathrm{e}^{-4\,\pi\,\sigma\,\gamma})\,\sigma\,\gamma}{8\,
\pi\,(1+ 5\,\gamma^2+4\,\gamma^4)}\,\l(\f{1}{-k\,\eta}\r)^4.\;\;
\end{eqnarray}
\end{subequations}
Upon using these expressions, we find that the squeezing amplitude~$r$ can 
be written as
\begin{equation}
\cosh\, (2\,\sr) \simeq \f{9\,\mathrm{e}^{-2\, \pi\, \sigma\,\gamma}
\sinh\,(2\, \pi\, \sigma\,\gamma)}{4\,\pi\,\sigma\,\gamma\,
(1+4\,\gamma^2)}\,\l(\f{1}{-k\,\eta}\r)^4.\label{eq:rk}
\end{equation}
This result implies that, towards the end of inflation, the squeezing amplitude 
$\sr$ behaves as (since $\sr$ is large) $\exp\,(2\,\sr)\propto a^4$ or, 
equivalently, $\sr \propto 2\,N$.
Actually, in the following sections, when we analyze the behavior of the squeezing 
amplitude in specific inflationary models, we shall see that such a behavior arises 
soon after the modes leave the Hubble radius.
The above result can be inverted to express the squeezing amplitude~$\sr$ (for 
large~$\sr$) as follows:
\begin{eqnarray}
\sr &\simeq & \ln \l(\f{3}{2}\r)-2\, \ln\l(\f{k}{\ke}\r)
-\pi\,\sigma\, \gamma\nn\\ 
& &+\,\f{1}{2}\ln\l[\f{\sinh\,(2\,\pi\, \sigma\,\gamma)}{\pi\,
\sigma\,\gamma\,(1+4\,\gamma^2)}\r],\label{eq:ra}
\end{eqnarray}
where $\ke$ represents the wave number that leaves the Hubble radius 
at the end of inflation. 
It is useful to note that, for small $\gamma$, we find that~$\sr$ behaves as
\begin{eqnarray}
\sr \simeq \ln\l(\f{3}{2}\r) + \f{1}{2}\,\ln\, 2 - 2\, \ln\l(\f{k}{\ke}\r) 
-\pi\, \sigma\, \gamma,\label{eq:r-gamma}
\end{eqnarray}
which suggests that~$\sr$ is linear in $\gamma$ in the limit.
Also, we had earlier pointed out that, for large~$\sr$, the quantum 
discord~$\delta$ behaves linearly with~$\sr$.
Later, when we evaluate the quantum discord $\delta$ numerically in the 
helical case, we shall find that, for small $\gamma$, the quantum discord 
depends linearly on the helicity parameter~$\gamma$.

To understand the behavior of the squeezing angle, we can make use of 
Eqs.~\eqref{eq:f2-g2-in-r} and write
\begin{equation}
\f{\langle \hat{\cP}^2\rangle}{k^2\,\langle \hat{\cA}^2\rangle}
=\f{1-\tanh\,(2\,\sr)\,\cos\,(2\,\varphi)}{1
+\tanh\,(2\,\sr)\,\cos\,(2\,\varphi)}
=\f{\vert g\vert^2}{k^2\,\vert f\vert^2}.
\end{equation}
At late times, when the squeezing amplitude~$\sr$ is large, $\tanh\,(2\,\sr)$ 
tends to unity, and the above relation simplifies to the following expression 
for the squeezing angle~$\varphi$:
\begin{equation}
\tan{\varphi} = \pm\f{\vert g\vert}{k\,\vert f\vert}.\label{eq:tanphi}
\end{equation}	
Upon using the solution~\eqref{eq:cA-h} in the de Sitter case, we find that, 
at late times, the squeezing angle reduces to
\begin{equation}
\tan{\varphi} \simeq -\sigma\,\gamma.\label{eq:tanphi-deS}
\end{equation}	
This implies that, while the angle~$\varphi$ vanishes for the non-helical 
modes, it is non-zero in the helical case and is of opposite signs for 
the two states of polarization.

Until now, we have focused on the $n=2$ case, which leads to a scale
invariant spectrum for the magnetic field. 
It is now interesting to examine if there can occur a situation (say, for 
a specific value of the parameter~$n$) wherein the squeezing amplitude over 
large scales is small. 
In other words, do there exist non-trivial coupling functions which lead 
to a small squeezing amplitude $\sr$ over large scales so that the modes 
remain close to the initial vacuum state at late times?
To understand this point, it proves to be helpful to express the squeezing
amplitude in terms of the power spectra of the electromagnetic fields.
Recall that the power spectra of the helical magnetic and electric fields,
say, $\pb(k)$ and $\pe(k)$, are defined as follows~\cite{Durrer:2010mq,
Anber:2006xt,Caprini:2014mja,Sharma:2018kgs}:
\begin{subequations}\label{eq:psbe-h}
\begin{eqnarray}
\pb(k) &=& \pb^{+}(k) +\pb^{-}(k)\nn\\
&=&\f{k^{5}}{4\,\pi^2\,a^{4}}\,
\l[\l\vert  \cA_k^{+}\r\vert^2 
+ \l\vert \cA_k^{-}\r\vert^2\r],\label{eq:psb-h}\\
\pe(k) &=& \pe^{+}(k) +\pe^{-}(k)\nn\\ 
&=&\! \f{k^3}{4\, \pi^2\, a^4}\, \l[\l\vert \cA_k^{+}{}'
- \f{J'}{J}\cA_k^+\r\vert^2+\l\vert \cA_k^{-}{}'
- \f{J'}{J}\cA_k^-\r\vert^2\r].\nn\\
\label{eq:pse-h}
\end{eqnarray}
\end{subequations}
Of course, in the non-helical case, the contributions from the two polarizations
to the power spectra become equal.
The above expressions for the power spectra and Eq.~\eqref{eq:r-g} suggest that 
we can express the squeezing amplitude~$r$ for a given polarization $\sigma$ as
follows:
\begin{equation}
\cosh\, (2\,\sr)=\f{4\,\pi^2\,a^4}{k^4}\,\l[\pb^{\sigma}(k)+\pe^{\sigma}(k)\r].
\end{equation}

Let us first consider the non-helical case.
For $n>1/2$, we find that, at late times, we can express the squeezing amplitude as 
\begin{equation}
\cosh\, (2\,\sr)\propto A_1\,  k^{-2\,n}+B_1\,k^{2-2\,n},
\end{equation}
whereas for $n<-1/2$, we have 
\begin{equation}
\cosh\, (2\,\sr)\propto A_2\, k^{2\,n+2}+ B_2\, k^{2\,n},
\end{equation}
where $(A_1,B_1,A_2,B_2)$ are constants~\cite{Tripathy:2021sfb,Tripathy:2022iev}.
Under either of these conditions, one of the two terms in the above expressions
dominates at small~$k$ suggesting a large squeezing amplitude.
In the helical case, for either polarization and for a non-zero~$n$, we have
\begin{equation}
\cosh\, (2\,\sr)\propto A_3\,k^{1-\vert 2\,n+1\vert} 
+ B_3\, k^{-2\,\vert n\vert},
\end{equation}
where $(A_3,B_3)$ are constants. 
Again, for any $n\neq 0$, one of the two terms dominates at small $k$ leading to
a significant squeezing amplitude.
The above discussion suggests that any non-trivial coupling function $J$ leaves
the large scale electromagnetic modes in a highly squeezed state at late times.


\subsection{In slow roll scenarios}

\begin{figure*}[t]
\includegraphics[width=0.325\linewidth]{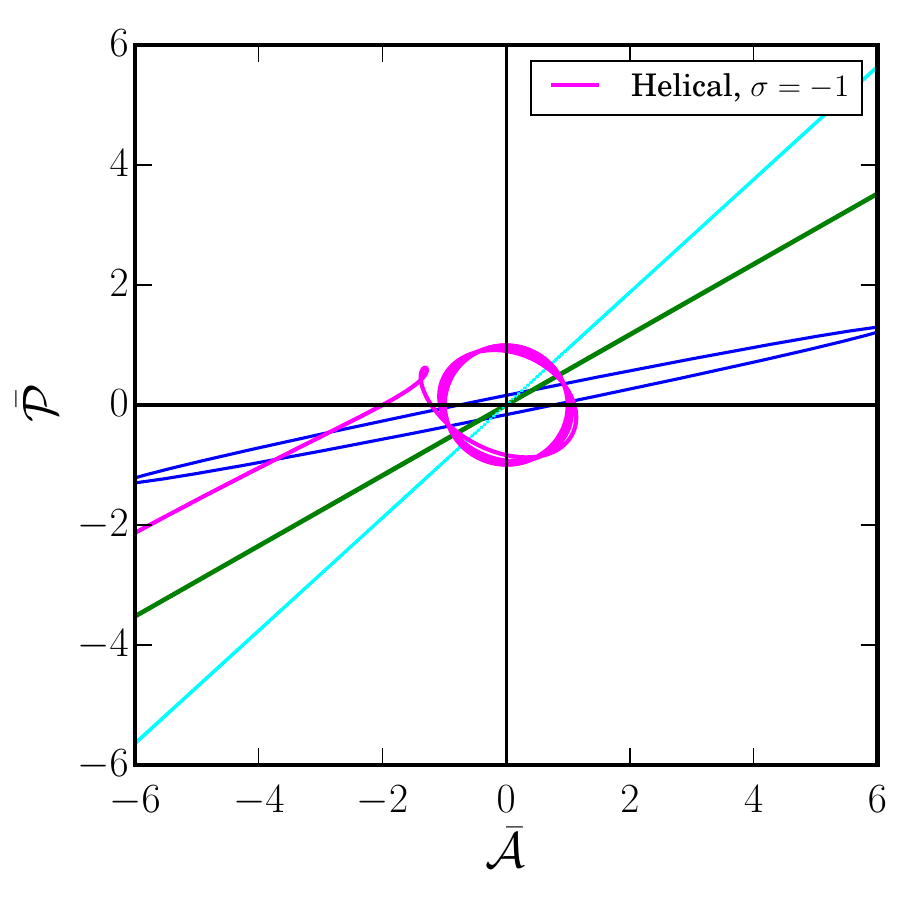}
\includegraphics[width=0.325\linewidth]{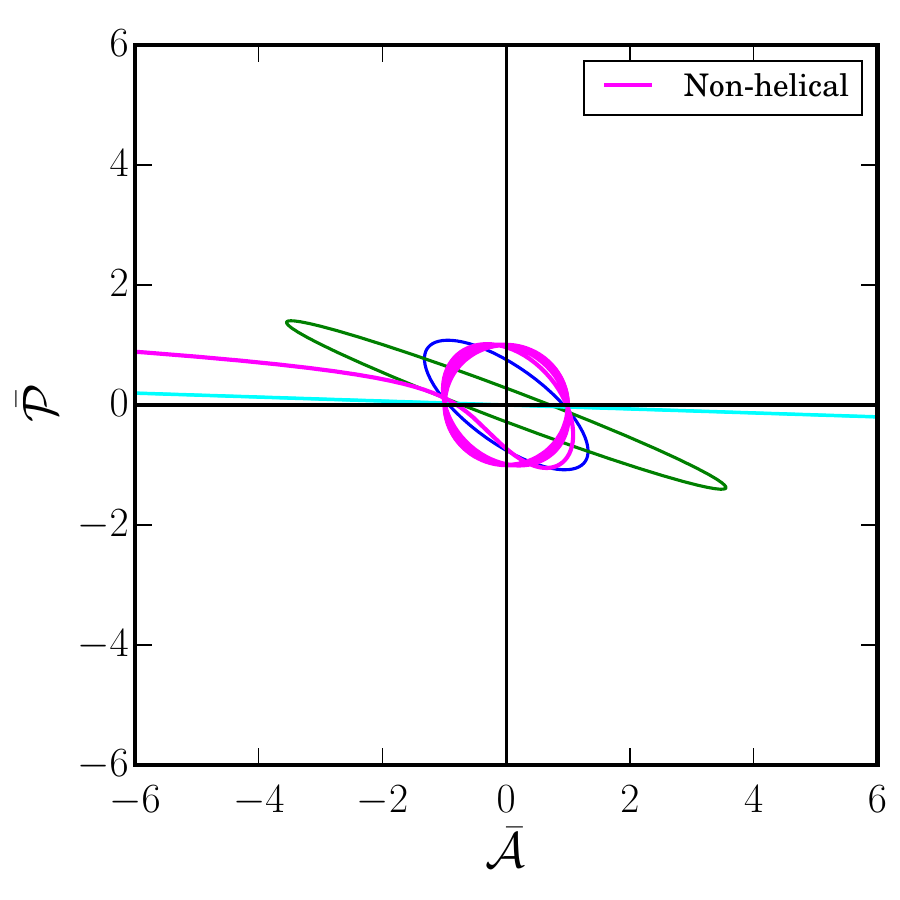}
\includegraphics[width=0.325\linewidth]{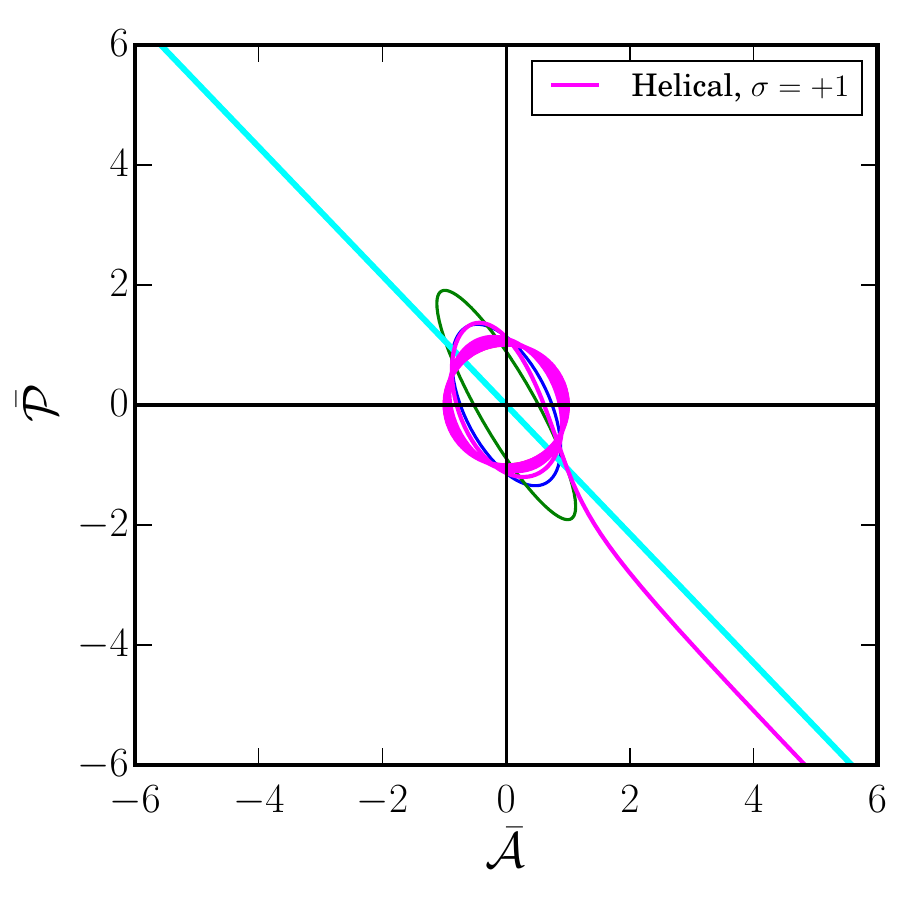}
\caption{We have plotted the evolution of the Wigner ellipse (in red,
blue, green and cyan) and the classical trajectory (in magenta) in the 
phase space $\bar{\cA}$-$\bar{\cP}$ for the electromagnetic mode in 
the Starobinsky model with the wave number corresponding to the CMB 
pivot scale, i.e. $k_\ast=0.05\, \mathrm{Mpc}^{-1}$.
We have plotted these quantities for the non-helical (in the middle)
as well as the helical cases (on the left and right for $\sigma=-1$ 
and~$\sigma=+1$, respectively).
The Wigner ellipses have been plotted at the following times:~when the
initial conditions are imposed on the mode in the sub-Hubble regime (in 
red, but hidden by the magenta curve), when $k=\sqrt{J''/J}$ (equivalent 
to the time of Hubble exit, in blue), on super-Hubble scales (in green) and 
closer to the end of inflation (in cyan).
We have set the helicity parameter $\gamma$ to be unity in plotting these 
figures. 
Clearly, the Wigner ellipse starts as a circle at early times and it 
is increasingly squeezed as time passes by.
In the non-helical case, the major axis of the ellipse eventually orients 
itself along the~$\bar{\cA}$ axis.
However, in the helical case, at late times, the major axis of the ellipse 
orients itself along a straight line with a non-vanishing slope.
As suggested by the condition~\eqref{eq:tanphi-deS}, while the helical
mode with the polarization state~$\sigma=-1$ has a positive slope, the 
state with~$\sigma=+1$ has a negative slope.
Moreover, as the relation~\eqref{eq:tanphi} suggests, we find that, at 
late times, the slope of the major axis of the ellipse is the same as 
that of the classical trajectory.}\label{fig:wes}
\end{figure*}
Let us now turn to understand the behavior of the Wigner ellipse, the
squeezing amplitude~$\sr$ and quantum discord~$\delta$ in specific 
inflationary models.
We shall first illustrate the behavior in slow roll inflation using the 
popular Starobinsky model.
The Starobinsky model is described by the potential
\begin{equation}
V(\phi)=V_0\,\l[1-\mathrm{exp}\l(-\sqrt{\f{2}{3}}\,\f{\phi}{\Mpl}\r)\r]^2,
\label{eq:sm1}
\end{equation}
where $V_0$ is a constant that is determined by the COBE normalization of
the scalar perturbations.
For $V_0 = 1.43 \times 10^{-10}\, \Mpl^4$, it is known that, at the pivot
scale of $k_\ast=5\times10^{-2}\, \mathrm{Mpc}^{-1}$ (often assumed to leave 
the Hubble radius about $N_\ast=50$ $e$-folds {\it before}\/ the end of inflation), 
the Starobinsky model leads to the scalar spectral index of $\ns=0.965$ and 
the tensor-to-scalar ratio of $r \simeq 4.3 \times 10^{-3}$, which fit 
the data from the anisotropies in the cosmic microwave background (CMB) 
very well~\cite{Planck:2018jri}. 
(The tensor-to-scalar ratio~$r$ should not be confused with the squeezing 
amplitude which is denoted in the same manner.)
In the slow roll approximation, the evolution of the field can be described 
in terms of the $e$-folds $N$ by the expression
\begin{eqnarray}
N-\Ne &\simeq & 
-\f{3}{4}\, \biggl[\mathrm{exp}\l({\sqrt{\f{2}{3}}}\,\f{\phi}{\Mpl}\r) 
- \mathrm{exp}\l({\sqrt{\f{2}{3}}}\,\f{\phie}{\Mpl}\r)\nn\\
& & -\, \sqrt{\f{2}{3}}\,\l(\f{\phi}{\Mpl} - \f{\phie}{\Mpl}\r)\biggr],
\end{eqnarray}
where $\phie$ is the value of the field at the $e$-fold $\Ne$ when inflation 
comes to an end.

As we mentioned, we require the non-conformal coupling function to behave 
as $J(\phi)\propto a^2$ in order to generate magnetic fields with a nearly
scale invariant spectrum.
Since the evolution of the field $\phi(N)$ differs from one model of inflation 
to another, to achieve $J(N) = \mathrm{exp}\,[2\,(N-\Ne)]$, the form of $J(\phi)$ 
will depend on the model at hand~\cite{Martin:2007ue,Tripathy:2021sfb,Tripathy:2022iev}.
In the Starobinsky model, we can choose the function $J(\phi)$ to be
\begin{eqnarray}
J(\phi) &=& \mathrm{exp}\,
\biggl\{-\f{3}{2}\, \biggl[\mathrm{exp}\l({\sqrt{\f{2}{3}}}\,
\f{\phi}{\Mpl}\r) - \mathrm{exp}\l({\sqrt{\f{2}{3}}}\,\f{\phie}{\Mpl}\r)\nn\\
& &-\, \sqrt{\f{2}{3}}\,\l(\f{\phi}{\Mpl} - \f{\phie}{\Mpl}\r)\biggr]\biggr\}.
\label{eq:J-sm1}
\end{eqnarray}
The equations governing the non-helical and helical electromagnetic modes 
corresponding to such a coupling function can be solved numerically to 
arrive at the power spectra of the magnetic and electric fields (in this 
regard, see Refs.~\cite{Tripathy:2021sfb,Tripathy:2022iev}).
We should point that the above coupling function $J(\phi)$ leads to minor 
deviations from the desired behavior of $J\propto a^2$ and, as a result, 
the spectrum of the magnetic field is {\it nearly}\/ scale invariant rather 
than a {\it strictly}\/ scale independent one~\cite{Tripathy:2021sfb}.

\begin{figure*}[t]
\centering
\includegraphics[width=0.495\linewidth]{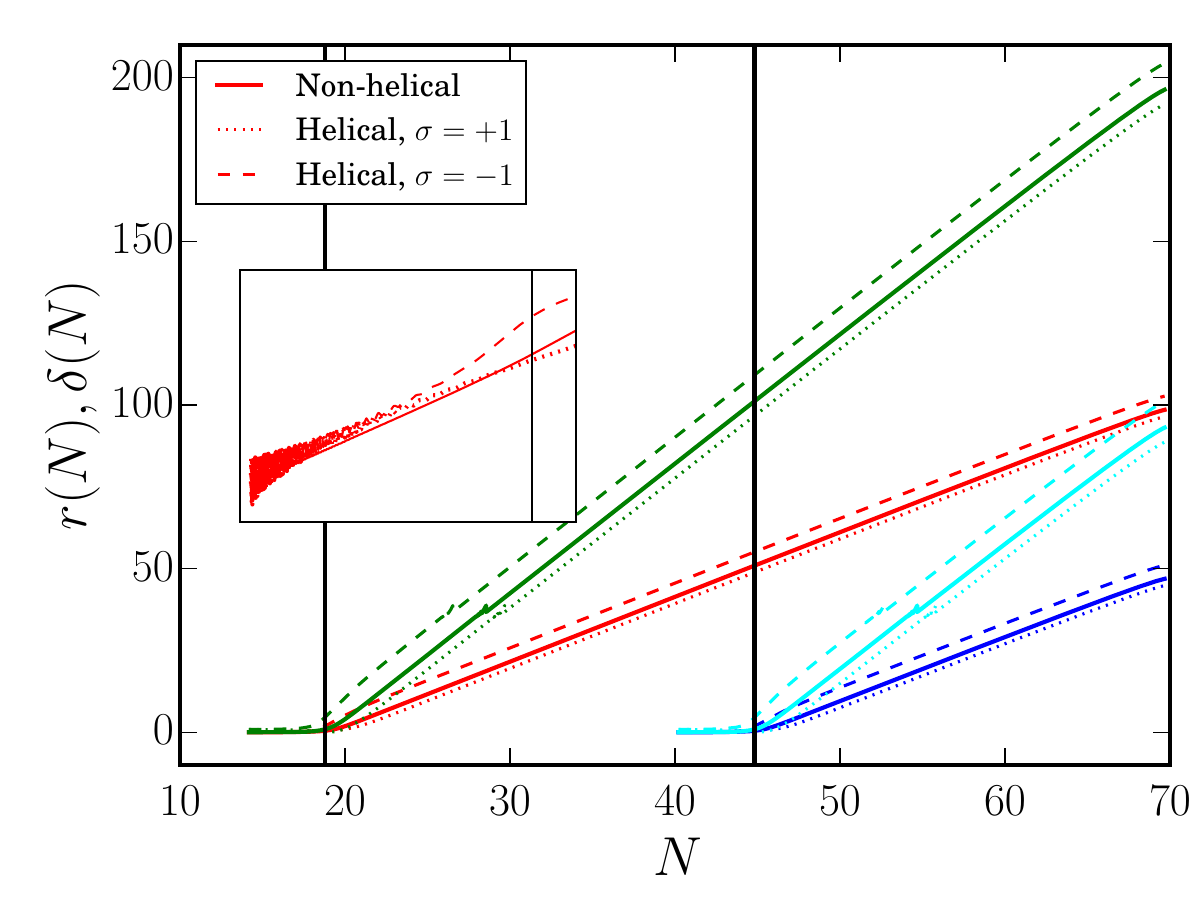}
\includegraphics[width=0.495\linewidth]{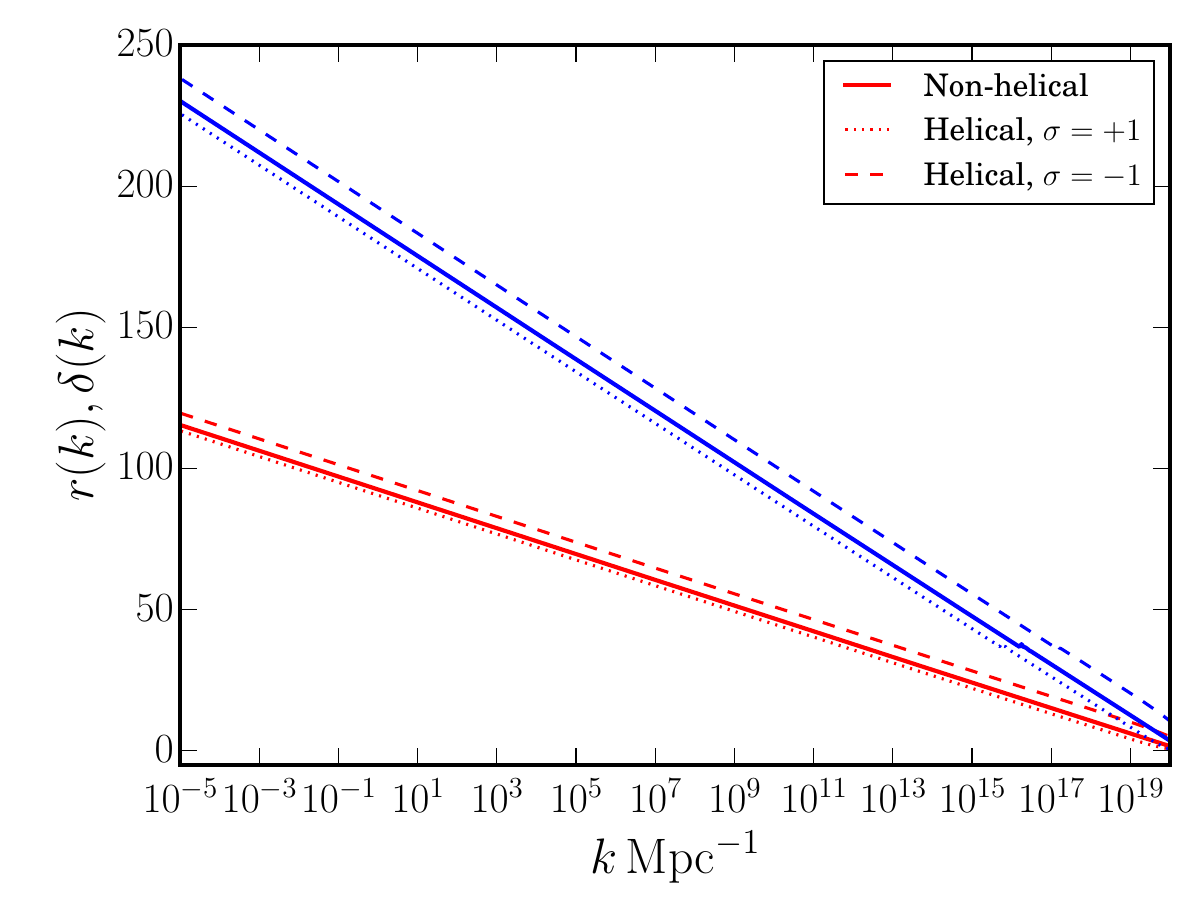}
\caption{The evolution of the squeezing amplitude~$\sr(N)$ (in red and blue) and 
quantum discord~$\delta(N)$ (in green and cyan) have been plotted (on the left)
for electromagnetic modes with two different wave numbers in the slow roll scenario
admitted by the Starobinsky model.
We have plotted the evolution for the CMB pivot scale of~$k_\ast =0.05\, 
\mathrm{Mpc}^{-1}$ (in red and green) and the small scale mode with the wave 
number $k = 10^{10}\,\mathrm{Mpc}^{-1}$ (in blue and cyan), which have been 
computed numerically.
The vertical lines (in black, on the left) indicate the time when $k^2 = J''/J$, 
i.e. roughly the time when the two modes leave the Hubble radius (at $N=18.75$ 
and $N=44.72$).
The inset (on the left, plotted on the log-linear scale) highlights the evolution 
of the squeezing amplitude associated with the pivot scale at early times.  
We have also plotted~(on the right) the `spectra' of the squeezing amplitude~$\sr(k)$
and the quantum discord~$\delta(k)$, evaluated at the end of inflation, for a wide
range of wave numbers.
Apart from the results for the non-helical case (which have been plotted as solid 
curves), we have plotted the results for the helical case (plotted as dotted and 
dashed lines, for $\sigma=+1$ and $\sigma=-1$, respectively).
We have set $\gamma=1$ in arriving at these figures.
As we have discussed in the text, the evolution of the squeezing amplitude and 
the quantum discord as well as their spectra behave in the manner expected from 
the analytical results in de Sitter inflation discussed earlier.}\label{fig:rdk-sr}
\end{figure*}
\begin{figure}
\centering
\includegraphics[width=1.0\linewidth]{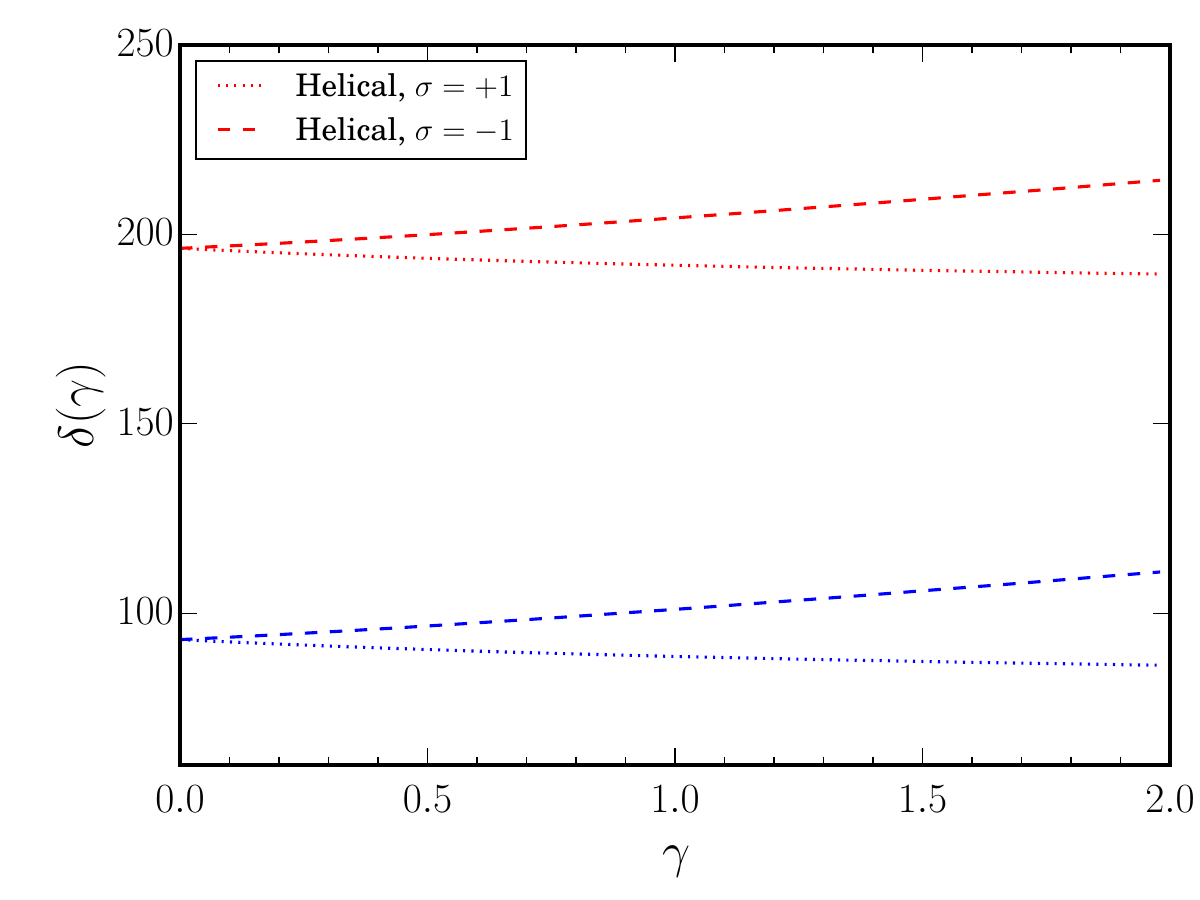}
\caption{The quantum discord $\delta$ evaluated at the end of inflation in
the case of the Starobinsky model has been plotted as a function of the 
helicity parameter~$\gamma$.
We have plotted the relation for two wave numbers, viz. the CMB pivot scale
$k_\ast = 0.05\, \mathrm{Mpc}^{-1}$ (in red) and $k=10^{10}\,
\mathrm{Mpc}^{-1}$ (in blue), for the two helical modes with $\sigma =+1$ 
(as dotted lines) and $\sigma=-1$ (as dashed lines). 
Note that, for small $\gamma$, $\delta$ behaves linearly with~$\gamma$.}
\label{fig:d-g}
\end{figure}
With the numerical solutions to the electromagnetic modes~$\cAs$ at hand, 
we can immediately evaluate the Wigner ellipse, the squeezing amplitude~$\sr$ 
and the quantum discord~$\delta$ using the expressions~\eqref{eq:phspace}
\eqref{eq:r-g} and~\eqref{eq:ee-ff} [with $y$ determined by Eq.~\eqref{eq:y}].
In Fig.~\ref{fig:wes}, we have illustrated the evolution of the Wigner ellipse
for a typical large scale mode in the Starobinsky model for the non-helical as
well as helical fields.
In the figure, we have also included the classical trajectory in phase space 
associated with the real parts of the solution~$f$ and the corresponding 
conjugate momentum~$g$ [cf. Eqs.~\eqref{eq:feq} and~\eqref{eq:g}] that 
determine the wave function~$\Psi(\cA,\eta)$.
In Fig.~\ref{fig:rdk-sr}, we have plotted the evolution of the 
quantities~$\sr$ and~$\delta$ as a function of the $e$-folds $N$ in the 
Starobinsky model for electromagnetic modes with two different wave numbers.
In the figure, we have also plotted the `spectra'~$\sr(k)$ and $\delta(k)$, 
i.e. the values of $\sr$ and $\delta$ evaluated at the end of inflation for 
a wide range of wave numbers.
In Fig.~\ref{fig:d-g}, we have plotted the dependence of the quantum
discord~$\delta$ on the helicity parameter~$\gamma$ for modes with the two 
different wave numbers.
The following points are clear from these figures.
Firstly, as expected, the Wigner ellipse starts as a circle and is 
increasingly squeezed with time.
Also, as suggested by Eq.~\eqref{eq:tanphi}, we find that, at late times, 
the slope of the major axis of the Wigner ellipse matches that of the 
classical trajectory.
Secondly, note that, on super-Hubble scales, the squeezing amplitude and the 
quantum discord associated with the $\sigma=-1$ helical modes have higher 
values when compared to the non-helical modes and the $\sigma=+1$ helical 
modes.
In fact, as should be clear from the inset in Fig.~\ref{fig:rdk-sr}, their 
values begin to differ even as they evolve in the sub-Hubble regime.
Thirdly, as expected from the results in the case of de Sitter inflation
discussed earlier, after the wave numbers have crossed the Hubble radius, 
$\sr(N)$ and $\delta(N)$ behave as $2\, N$ and $4\,N$, respectively, in 
all the cases.
Fourthly, in the linear-log plot, the spectra $\sr(k)$ and $\delta(k)$ of 
the squeezing amplitude and quantum discord behave as $(k/\ke)^{-2}$ and 
$(k/\ke)^{-4}$, as we had discussed [cf. Eq.~\eqref{eq:ra}].
Lastly, it is clear from Fig.~\ref{fig:d-g} that quantum discord $\delta$ 
behaves linearly with the helicity parameter for small~$\gamma$
[cf.~Eq.~\eqref{eq:r-gamma}].


\subsection{In scenarios involving departures from slow roll}

Let us now turn to understand the behavior of the squeezing amplitude~$\sr$ 
and quantum discord~$\delta$ in situations involving departures from slow 
roll.
It is well known that specific features in the inflationary scalar power 
spectrum improve the fit to the CMB data, when compared to the nearly scale 
invariant power spectra that arise in slow roll scenarios (for a partial
list of efforts in this regard, see Refs.~\cite{Contaldi:2003zv,Powell:2006yg,
Jain:2008dw,Jain:2009pm,Hazra:2010ve,Aich:2011qv,Benetti:2013cja,Hazra:2014jka,
Hazra:2014goa,Chen:2016zuu,Chen:2016vvw,Ragavendra:2020old,Antony:2021bgp}).
Moreover, recently, there has been a considerable interest in the literature 
to study inflationary models that generate enhanced power on small scales and
lead to the formation of a significant number of primordial black holes 
(PBHs) (in this context, see, for instance, Refs.~\cite{Garcia-Bellido:2017mdw,
Ballesteros:2017fsr,Germani:2017bcs,Dalianis:2018frf,Bhaumik:2019tvl,
Ashoorioon:2019xqc,Ragavendra:2020sop,Dalianis:2020cla,Ragavendra:2023ret}).
If such features are to be generated, then the inflationary potential should
admit deviations from slow roll. 
In fact, the stronger the feature in the scalar power spectrum (as is, say, 
required to produce a considerable number of PBHs), the sharper should be 
the departures from slow roll inflation.
Interestingly, in a recent work, we have illustrated that such deviations from 
slow roll inflation also lead to strong features in the spectra of magnetic 
fields~\cite{Tripathy:2021sfb}
We have also shown that, while it is possible to restore scale invariance
of the spectrum of the magnetic field in some situations, it is achieved at 
the cost of severe fine tuning~\cite{Tripathy:2022iev}.
In this section, we shall discuss the behavior of the squeezing amplitude 
and quantum discord in single and two field models of inflation that permit 
{\it strong}\/ departures from slow roll. 


\subsubsection{Single field models}

We shall first consider two single field models that lead to sharp 
departures from slow roll inflation and hence to strong features 
in the scalar power spectra.
The first model we shall consider is described by the 
potential~\cite{Jain:2008dw,Jain:2009pm,Ragavendra:2020old}
\begin{equation}
V(\phi)=\f{m^2}{2}\, \phi^2-\f{2\,m^2}{3\,\phi_0}\, \phi^3
+\f{m^2}{4\, \phi_0^2}\,\phi^4,\label{eq:pi1}
\end{equation}
and we shall work with the following values of the two parameters 
involved:~$m=7.17\times10^{-8}\,\Mpl$ and $\phi_0=1.9654\,\Mpl$.
Also, we shall choose the initial values of the field and the first slow 
roll parameter to be $\phii =12.0\,\Mpl$ and $\e1i= 2 \times 10^{-3}$. 
For these values of parameters and initial conditions, inflation lasts 
for about $110$~$e$-folds in the model, which is much longer than the 
duration typically considered.
However, if we assume that the pivot scale exits the Hubble radius
about $91$~e-folds before the termination of inflation, we find that
the model leads to a suppression in the scalar power spectrum on the 
largest scales and thereby to a moderate improvement in the fit to 
the CMB data (in this context, see Ref.~\cite{Ragavendra:2020old}).
The second model that we shall consider is described by the 
potential~\cite{Dalianis:2018frf}
\begin{eqnarray}
V(\phi) 
&=& V_0\,\biggl\{\mathrm{tanh}\l(\f{\phi}{\sqrt{6}\,\Mpl}\r)\nn\\ 
& &+\, A\,\sin\l[\f{1}{f_\phi}\,
\mathrm{tanh}\l(\f{\phi}{\sqrt{6}\,\Mpl}\r)\r]\biggr\}^2.\label{eq:usr}
\end{eqnarray}
We shall choose to work with the following values of the parameters:~$V_0 
= 2\times10^{-10}\,\Mpl^4$, $A = 0.130383$ and $f_\phi 
= 0.129576$.
We find that, if we set the initial value of the field to be $\phii=6.1\,\Mpl$, 
with $\e1i=10^{-4}$, we obtain about $66$~e-folds of inflation in the model.
Moreover, we shall assume that the pivot scale exits the Hubble radius about
$56.2$~e-folds prior to the termination of inflation.
This model generates enhanced power on small scales which results in the
production of a significant number of PBHs.
Both these models contain a point of inflection.
It is located at $\phi_0=1.9654\,\Mpl$ in the first model and at $\phi_0 
= 1.05\,\Mpl$ in the second~\cite{Ragavendra:2020old,Ragavendra:2020sop}.
The point of inflection leads to an epoch of ultra slow roll inflation
which is responsible for the sharp features in the power spectra (for 
a detailed discussion in this regard, see the recent 
review~\cite{Ragavendra:2023ret}).

\begin{figure*}
\includegraphics[width=0.495\linewidth]{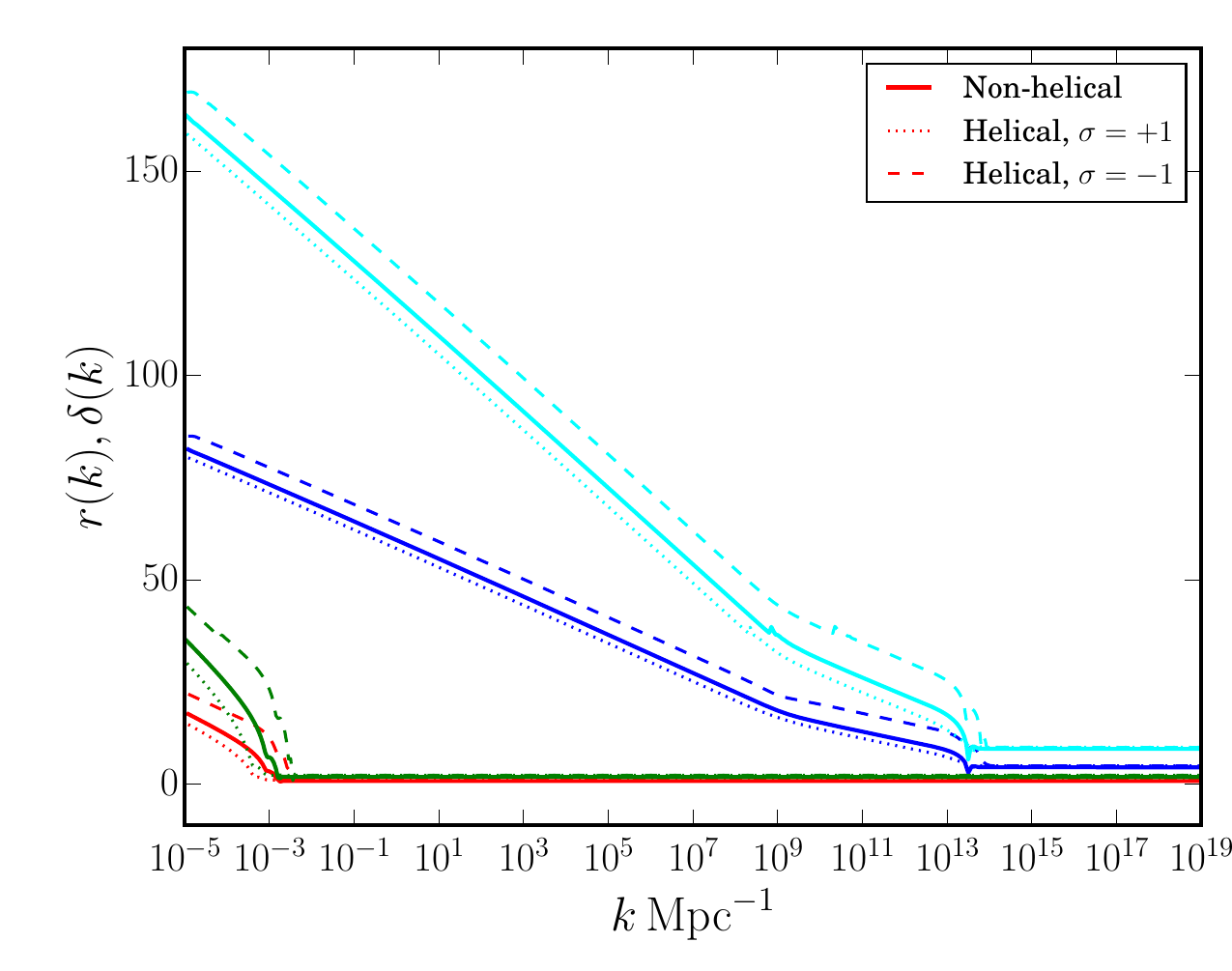}
\includegraphics[width=0.495\linewidth]{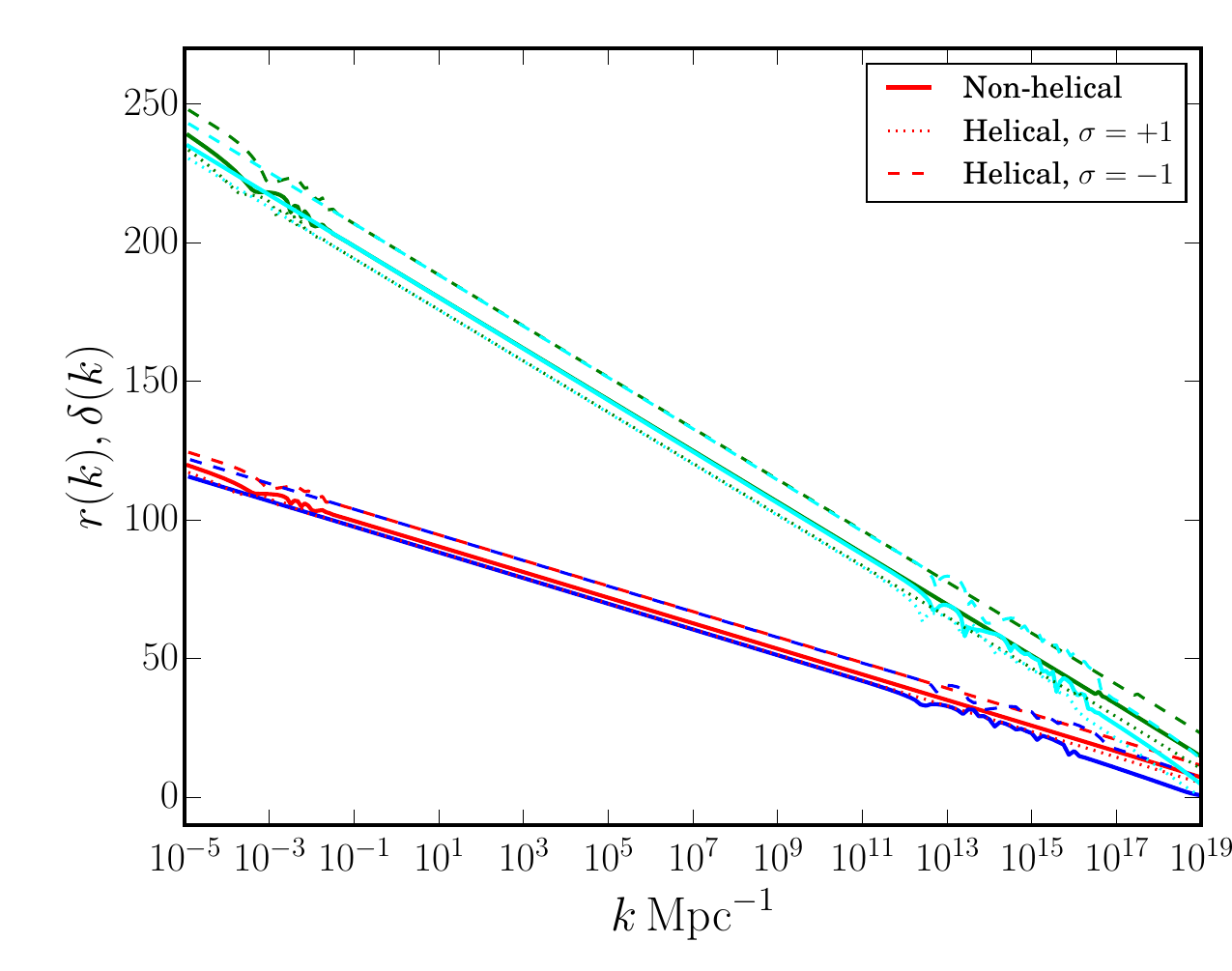}
\caption{The `spectra' of the squeezing amplitude~$\sr(k)$ (in red and blue) 
and quantum discord~$\delta(k)$ (in green and cyan) that arise in the single 
and two field inflationary models of our interest have been plotted (on the 
left and right, respectively) for a wide range of wave numbers.
Note that, in the case of single field models, $\sr(k)$ and $\delta(k)$ (plotted 
on the left) is rather small over wave numbers that leave the Hubble radius 
after the onset of ultra slow roll.
However, in the case of the two field models, the quantities $\sr(k)$ and $\delta(k)$
behave virtually in the same manner as they do in the slow roll scenario.
This behavior has been achieved with the aid of the additional field 
available in the two field models.}\label{fig:rdk-usr-tfm} 
\end{figure*}
Due to the strong departures from slow roll, in general, it proves to be 
challenging to arrive at analytical solutions for the background scalar 
field in these models.
As we discussed, to arrive at a nearly scale invariant spectrum for the
magnetic field, we need to choose the non-conformal coupling function to 
behave as $J\propto a^2$.
Since there does not exist analytical solutions for the scalar field in 
the models of our interest, we are unable to construct an analytical form 
for $J(\phi)$ that leads to the desired behavior, as we had done in the 
case of the Starobinsky model [cf. Eq.~\eqref{eq:J-sm1}].
Therefore, we have to resort to a numerical approach to arrive at a 
suitable non-conformal coupling function~$J(\phi)$ (for discussions 
in this regard, see our recent efforts~\cite{Tripathy:2021sfb,
Tripathy:2022iev}).
However, because of the points of inflection, in these potentials, the
non-conformal coupling function $J(\phi)$ hardly evolves as the field
approaches the point of inflection and the epoch of ultra slow roll 
sets in.
Such a behavior generates magnetic fields with spectra that have a 
strong scale dependence.
The resulting spectra of the magnetic field are scale invariant over 
large scales (i.e. over wave numbers that leave the Hubble radius prior 
to the onset of the epoch of ultra slow roll) and behave as $k^4$ on 
small scales (i.e. over wave numbers that leave the radius after ultra 
slow roll has set in).
Moreover, it is found that the scale invariant amplitude of the magnetic 
field on large scales is strongly suppressed, with the amplitude being 
lower when the onset of ultra slow roll is earlier.
In Fig.~\ref{fig:rdk-usr-tfm}, we have plotted the spectra of the squeezing
amplitude~$\sr(k)$ and quantum discord~$\delta(k)$ for the magnetic fields
generated in the two inflationary models described above.
Note that, on larger scales (corresponding to the scale invariant
domain in the spectra of the magnetic field), the quantities $\sr(k)$ and
$\delta(k)$ behave as in the slow roll case.
However, over smaller scales wherein the spectra of the magnetic field
behave as $k^4$, we find that $\sr(k)$ and $\delta(k)$ are rather small
suggesting that the modes have not evolved significantly from the 
Bunch-Davies vacuum.


\subsubsection{In two field models}

We had pointed out above that, in the case of single field models permitting
an epoch of ultra slow roll, the non-conformal coupling function $J(\phi)$
hardly evolves after the onset of ultra slow roll.
Such a behavior leads to magnetic field spectra which have strongly suppressed 
scale invariant amplitudes on larger scales and~$k^4$ dependence on smaller 
scales.
We have recently shown that these challenges can be circumvented in two
field models of inflation~\cite{Tripathy:2022iev}.
In models involving two fields, it is possible to construct inflationary 
scenarios that generate sharp features in the scalar power spectra (on either 
large or small scales) and design suitable non-conformal coupling functions
that lead to nearly scale invariant spectra of magnetic fields with the 
desired amplitudes.
We shall now discuss the behavior of the squeezing amplitude $\sr$ and the 
quantum discord $\delta$ in such models.

We shall consider two models, one of which leads to a suppression in the 
spectrum of curvature perturbations on large scales and another which 
leads to an enhancement in the scalar power on small scales.
The two field models are described by the action~\cite{Braglia:2020fms,
Braglia:2020eai}
\begin{eqnarray}
S[\phi,\chi] &=& \int \d^4x\, \sqrt{-g}\,
\biggl[-\f{1}{2}\,\pa_\mu{\phi}\, \pa^\mu {\phi}\nn\\
& & -\, \f{f(\phi)}{2}\,\pa_\mu\chi\, \pa^\mu {\chi}-V(\phi,\chi)\biggr],
\end{eqnarray}
where, evidently, while $\phi$ is a canonical scalar field, $\chi$ is a 
non-canonical scalar field due to the presence of the function $f(\phi)$ 
in the term describing its kinetic energy. 
We shall assume that $f(\phi)= \mathrm{exp}\,(2\,\bar{b}\,\phi)$, where 
$\bar{b}$ is a constant.
The first model we shall consider is described by the 
potential~\cite{Braglia:2020fms}
\begin{equation}
V(\phi,\chi)=\f{m_\phi^2}{2}\,\phi^2 
+ V_0\,\f{\chi^2}{\chi_0^2+\chi^2}.\label{eq:sup-ls}
\end{equation}
We shall work with the following values of the parameters 
involved:~$(m_\phi/\Mpl,V_0/\Mpl^4,\chi_0/\Mpl,\bar{b}\,\Mpl)
=(1.672 \times10^{-5},2.6\times10^{-10}, \sqrt{3},1.0)$.
Also, we shall choose the initial values of the fields to 
be $\phi_\mathrm{i}= 8.8\,\Mpl$, $\chi_\mathrm{i} =5.76\, 
\Mpl$, and set $\epsilon_{1\mathrm{i}}=2.47 \times 10^{-2}$.
For these values of the parameters, we obtain about $78$ $e$-folds 
of inflation.
The second potential we shall investigate can be arrived at by 
interchanging the individual potentials for the two fields we 
considered above, and is given by~\cite{Braglia:2020eai}
\begin{equation}
V(\phi,\chi)=V_0\,\f{\phi^2}{\phi_0^2+\phi^2}
+\f{m_\chi^2}{2}\,\chi^2.\label{eq:ss-peak}
\end{equation}
We shall work with the following  values of the 
parameters:~$(V_0/\Mpl^4,\phi_0/\Mpl,m_\chi/\Mpl,\bar{b}\,\Mpl)
=(7.1 \times10^{-10}, \sqrt{6}, 1.19164\times10^{-6},7.0)$ and 
assume that $\phi_\mathrm{i}=7.0\,\Mpl$, $\chi_\mathrm{i}=7.31\,\Mpl$
and $\e1i=4.32 \times 10^{-4}$.
For these parameters and initial conditions, we obtain about 
$84$~$e$-folds of inflation in the model.

These models lead to two stages of slow roll inflation, with each
stage being driven by one of the two fields.
There arises a sharp turn in the field space as the transition 
from one stage to another occurs.
The transition leads to a tachyonic instability and the isocurvature
perturbations source the curvature perturbations associated with 
wave numbers which leave the Hubble radius during the turn in the 
field space~\cite{Braglia:2020eai,Braglia:2020fms}.
The first of the above two models leads to a suppression in scalar
power on large scales~\cite{Braglia:2020eai}, while the second leads 
to an enhancement in power on small scales~\cite{Braglia:2020fms}.
In these models, we can construct non-conformal coupling functions 
that depend on the field driving slow roll inflation in each of the 
two stages.
The two functions can then be combined together to arrive at a 
complete non-conformal coupling function~$J(\phi,\chi)$ that 
largely leads to the desired behavior of $J\propto a^2$, barring 
the period around the transition in the dependence of $J$ on one 
field to the other (for a detailed discussion in this regard, see
Ref.~\cite{Tripathy:2022iev}).
We should clarify that the non-conformal coupling function has to be 
fine tuned to a certain extent to avoid substantial deviations from 
the $J \propto a^2$ behavior around the transition. 
The resulting non-conformal coupling function leads to a nearly
scale invariant spectrum for the magnetic field which contains 
some features around the range of wave numbers which leave the 
Hubble radius close to the time of the transition.
In Fig.~\ref{fig:rdk-usr-tfm}, we have plotted the spectra of
the squeezing amplitude $\sr(k)$ and quantum discord $\delta(k)$
that arise in the two field models that we have introduced above.
Clearly, the quantities $\sr(k)$ and $\delta(k)$ contain some small
wiggles around the domain (in wave numbers) when the spectra of the 
magnetic field contain features (for the spectra of the magnetic
field that arise in these cases, see Ref.~\cite{Tripathy:2022iev}).
Otherwise, the spectra of the squeezing amplitude and quantum discord 
behave in the same manner as in the slow roll scenario (cf. Fig.~\ref{fig:rdk-sr}).


\section{Discussion}\label{sec:s}

In this work, we have examined the evolution of the quantum state of the 
non-helical as well as helical electromagnetic fields generated during
inflation.
We have tracked the evolution of the state of the electromagnetic field
using measures such as the Wigner ellipse, squeezing amplitude and 
quantum discord.
We find that, in a manner similar to the case of the scalar perturbations, 
the squeezing amplitude and quantum discord associated with the 
non-helical electromagnetic modes evolve linearly with $e$-folds on 
super-Hubble scales.
Interestingly, in case of the helical electromagnetic field, the squeezing 
amplitude as well as the quantum discord of one of the two states of 
polarization ($\sigma=-1$) is enhanced when compared to the non-helical case, 
whereas they are suppressed for the other (i.e. for the polarization state with
$\sigma=+1$).
In fact, the enhancement (or suppression) occurs as the helical modes leave
the Hubble radius and, on super-Hubble scales, the squeezing amplitude (and
quantum discord) behave as a function of $e$-folds just as in the non-helical 
case.
We find that, over the range of the values of the helicity parameter $\gamma$
that we have considered, the enhancement of the squeezing amplitude and
quantum discord is not very significant.
We had limited ourselves to $\gamma \lesssim 2.5$ to avoid the issue of
backreaction due to the helical modes on the background (in this context,
see Ref.~\cite{Tripathy:2021sfb}).

On a related note, we find that there is a similarity between the 
effects on the electromagnetic field due to the violation of parity during 
inflation that we have discussed and the Schwinger effect associated with, 
say, a charged scalar field that arises when a constant electric field is 
present in the de Sitter spacetime (for relatively recent discussions on 
the Schwinger effect in the de Sitter spacetime, see, for instance, 
Refs.~\cite{Frob:2014zka,Kobayashi:2014zza,Sharma:2017ivh,Banyeres:2018aax}).
The electric field provides a direction breaking the isotropy of the
FLRW universe.
As a result, the modes of the charged field behave in a fashion akin to the 
helical electromagnetic field with the modes propagating along the direction 
of the electric field behaving differently from the modes travelling in the
opposite direction.
We have discussed these points in some detail in App.~\ref{app:se}.

Let us conclude by highlighting a few different directions in which further
investigations need to be carried out.
Firstly, it will be worthwhile to examine carefully~(including backreaction) 
the effects due to helicity for larger values of $\gamma$~\cite{Durrer:2023rhc}.
Secondly, due to a technical difficulty we had faced in the helical case (as 
described in Sec.~\ref{sec:qd}), we have worked with two different conjugate 
momenta [given by Eqs.~\eqref{eq:cm} and~\eqref{eq:cm1}] while calculating the 
squeezing amplitude~$\sr$ and the quantum discord~$\delta$.
While we have been able to evaluate the squeezing amplitude~$\sr$ with the wave 
function starting in the Bunch-Davies vacuum (as described in Sec.~\ref{sec:sv1}),
we have evaluated the quantum discord~$\delta$ with the wave function beginning
in a slightly squeezed initial state (arising due to the choice of the conjugate
momentum; in this context, see the discussion in App.~\ref{app:cm}).
We need to overcome this challenge and evaluate the quantum discord of the 
system associated with the action~\eqref{eq:Af-k-k-1}.
Thirdly, the effects due to parity violation we have encountered can also 
occur in the case of tensor perturbations described by modified theories 
such as the Chern-Simons theory of gravitation (in this regard, see, for 
example, Ref.~\cite{Peng:2022ttg}).
Lastly, it has been pointed out that the quantum-to-classical transition of 
the primordial scalar perturbations can affect the extent of non-Gaussianities
generated in the early universe~\cite{DaddiHammou:2022itk}.
It will be interesting to consider the effects that arise due to the decoherence
of the scalar~(or the tensor) perturbations and the magnetic fields on the 
cross-correlations between the magnetic fields and the curvature (or the tensor) 
perturbations~\cite{Jain:2012ga,Jain:2012vm,Motta:2012rn,
Chowdhury:2018blx,Chowdhury:2018mhj,Jain:2021pve}.
We are presently investigating some of these issues.


\acknowledgements

The authors wish to thank J\'{e}r\^{o}me Martin for discussions and comments 
on the manuscript.
ST would like to thank the Indian Institute of Technology Madras, Chennai, 
India, for support through the Half-Time Research Assistantship. 
RNR is supported through a postdoctoral fellowship by the Indian Association 
for the Cultivation of Science, Kolkata, India.
LS wishes to acknowledge support from the Science and Engineering Research 
Board, Department of Science and Technology, Government of India, through 
the Core Research Grant~CRG/2020/003664. 


\appendix


\section{On the choice of conjugate momentum}\label{app:cm}

Recall that, initially, we had arrived at the action~\eqref{eq:Af} to describe 
the Fourier modes of the helical electromagnetic field.
If we focus on the electromagnetic mode associated with a single wave 
number (as we did in Secs.~\ref{sec:sv1}, \ref{sec:we1} and~\ref{sec:sp1}),
the fiducial variable $\cA$---which stands for either~$\cA^{\sigma}_{\bmk\mathrm{R}}$ 
or~$\cA^{\sigma}_{\bmk\mathrm{I}}$ [introduced in Eq.~\eqref{eq:ARI}]---is 
described by the following Lagrangian density in Fourier space:
\begin{equation}
\mathcal{L} = \f{1}{2}\,\cA'{}^2-\kappa\,\cA'\,\cA-\f{\mu^2}{2}\,\cA^2.
\label{eq:LcA} 
\end{equation}
In this appendix, we shall explain the reason for adding the total time 
derivative~\eqref{eq:TD} to the original action~\eqref{eq:Af-in} to 
arrive at the modified action~\eqref{eq:Af} or, equivalently, the 
Lagrangian~\eqref{eq:Lptd-f1} for the variable~$\cA$. 


\subsection{Choices of momenta and initial conditions}

Note that the conjugate momentum associated with the original 
Lagrangian~\eqref{eq:LcA} is given by
\begin{eqnarray}
\cP=\cA'-\kappa\,\cA.\label{eq:cm1}
\end{eqnarray}
The corresponding Hamiltonian can be obtained to be
\begin{equation}
\mathcal{H}= \f{\cP^2}{2}+\kappa\,\cP\,\cA
+\f{\widetilde{\omega}^2}{2}\,\cA^2,
\end{equation}
where the quantity~$\widetilde{\omega}^2$ is given Eq.~\eqref{eq:omegatilde2}.
The \Sch equation governing the wave function~$\Psi(\cA,\eta)$ 
corresponding to the above Hamiltonian is given by
\begin{equation}
i\,\f{\pa \Psi}{\pa \eta}
=-\f{1}{2}\,\f{\pa^2\Psi}{\pa \cA^2}
-\f{i\,\kappa}{2}\,\l(\Psi+2\,\cA\,\f{\pa \Psi}{\pa \cA}\r)
+\f{\widetilde{\omega}^2}{2}\,\cA^2\,\Psi.\quad
\end{equation}
Upon using the Gaussian ansatz~\eqref{eq:wfncA} for the wave 
function~$\Psi(\cA,\eta)$ in this \Sch equation, we obtain that
\begin{equation}
\Omega' = -i\,\Omega^2-2\,\kappa\,\Omega+i\,\widetilde{\omega}^2. 
\label{eq:omega1}
\end{equation}

If we now use the definition~\eqref{eq:domega} of $\Omega$ in 
the above equation, but with $g$ being given by
\begin{equation}
g=f'-\kappa\,f,\label{eq:g1}    
\end{equation}
then we arrive at the same equation for~$f^\ast$ that we had 
obtained earlier, viz. Eq.~\eqref{eq:feq}.
This should not come as a surprise since, classically, Lagrangians that 
differ by a total time derivative lead to the same equation of motion.
Moreover, since the wave function is assumed to be of the same form, 
we obtain the same Wigner function as had obtained before, i.e. as in 
Eq.~\eqref{eq:wfn-ga}, but with $\cP$ being the new conjugate momentum
defined in Eq.~\eqref{eq:cm1}.
However, for $I =J$ and $J \propto \eta^{-n}$, we find that, at early 
times, while $f$ behaves as in Eq.~\eqref{eq:fnh}, $g$ behaves as
\begin{equation}
g=f'-\kappa\,f \simeq -i\,\sqrt{\f{k}{2}}\, 
(1+i\,\sigma\,\gamma)\,\mathrm{e}^{-i\,k\,\eta}.\label{eq:gnh1}
\end{equation}
These~$f$ and~$g$ lead to the same Wronskian~\eqref{eq:wronskian} 
that we had obtained earlier. 
Also, in such a case, we have $\Omegar=k$ and $\Omegai=\sigma\, \gamma\, k$,
resulting in the following condition for the Wigner ellipse:
\begin{equation}
\bar{\cA}^2+\l(\cP+\sigma\,\gamma\,\bar{\cA}\r)^2=1.  
\end{equation}
Moreover, for the above initial conditions on~$f$ and~$g$, from 
Eqs.~\eqref{eq:rphi-g}, we obtain that $\mathrm{cosh}\,(2\,\sr)
=1+(\gamma^2/2)$, while $\mathrm{cos}\,(2\,\varphi)
=\pm \gamma/\sqrt{4+\gamma^2}$.
These imply that, when $\gamma$ is non-zero (i.e. in the helical case), 
at early times, the Wigner ellipse is not a circle.
It starts as an ellipse with its major axis oriented at the angle~$\varphi$ 
with respect to the~$\bar{\cA}$ axis.

If we now instead add a different total time derivative to the 
original Lagrangian~\eqref{eq:LcA} as follows
\begin{equation}
\mathcal{L}= \f{1}{2}\,\cA'{}^2-\kappa\,\cA'\,\cA-\f{\mu^2}{2}\,\cA^2
+\f{\d}{\d \eta}\l(\f{1}{2}\kappa\,\cA^2\r),\quad
\end{equation}
then it simplifies to the form
\begin{equation}
\mathcal{L} = \f{1}{2}\,\cA'{}^2-\f{1}{2}\,\omega^2\,\cA^2\label{eq:Lptd-f2}
\end{equation}
with $\omega^2$ being given by Eq.~\eqref{eq:omega2}.
The corresponding conjugate momentum is given by
\begin{equation}
\cP=\cA'\label{eq:cm2}
\end{equation}
and the associated Hamiltonian can be immediately obtained to be
\begin{equation}
\mathcal{H}=\f{1}{2}\,\cP^2+\f{1}{2}\,\omega^2\,\cA^2.
\end{equation}
The \Sch equation describing the wave function $\Psi(\cA,\eta)$ in
such a case is given by
\begin{equation}
i\,\f{\pa \Psi}{\pa \eta}
=-\f{1}{2}\,\f{\pa^2\Psi}{\pa \cA^2}+\,\f{1}{2}\,\omega^2\,\cA^2\,\Psi.
\end{equation}
The Gaussian ansatz~\eqref{eq:wfncA} for the wave function leads to
the following equation for $\Omega$:
\begin{equation}
\Omega' = -i\,\Omega^2+i\,\omega^2.
\end{equation}
Upon substituting the definition~\eqref{eq:domega} of $\Omega$ in this 
differential equation, but with 
\begin{equation}
g=f',\label{eq:gnh-2}
\end{equation} 
then we obtain Eq.~\eqref{eq:feq} for $f^\ast$, as one would have expected.
Note that, in such a situation, as with the Lagrangian~\eqref{eq:Lptd-f1}, 
at early times, $f$ and~$g$ reduce to the forms in Eqs.~\eqref{eq:fnh}
and~\eqref{eq:gnh} implying that the initial Wigner ellipse is a circle.
Also, as in the original case, we have, at early times,  
$\mathrm{cosh}\,(2\,\sr)=1$, while $\mathrm{cos}\,(2\,\varphi)$ is
undetermined.

In order to ensure that the system starts in the standard Bunch-Davies 
vacuum at early times with no squeezing involved, we have worked with 
the modified Lagrangian~\eqref{eq:Lptd-f1} instead of the 
Lagrangian~\eqref{eq:LcA}.
As we have discussed earlier, if we work in terms of the corresponding
conjugate momentum~$\cP$ [cf. Eq.~\eqref{eq:cm}], we obtain a Wigner 
ellipse which starts as a circle at early times, as desired [cf. 
Eq.~\eqref{eq:we-bd}].
But, such a behavior of the Wigner ellipse and the squeezing parameters
at early times is also encountered when the system is described by the 
Lagrangian~\eqref{eq:Lptd-f2}], as we discussed above.
Could we have also worked with the Lagrangian~\eqref{eq:Lptd-f2}?
It seems that the Lagrangian~\eqref{eq:Lptd-f1} is an appropriate choice. 
Let us illustrate this point with a simple example.


\subsection{A simple example}

To illustrate our point, we shall focus on the non-helical electromagnetic 
field (i.e. when $\gamma=0$) and consider the case wherein $J= (\eta/\ee)^{-n}$.
The trivial case, of course, corresponds to the conformally coupled field 
wherein $n=0$.
In such a case, all the momenta $\cP$ we have encountered, i.e. those given 
by Eqs.~\eqref{eq:cm}, \eqref{eq:cm1} and~\eqref{eq:cm2}, turn out to be 
the same and the quantities $f$ and $g$ are {\it exactly}\/ given by 
Eqs.~\eqref{eq:fnh} and~\eqref{eq:gnh} {\it at all times}.\/
Therefore, $\mathrm{cosh}\,(2\,\sr)=1$ forever, while $\mathrm{cos}\,
(2\,\varphi)$ remains undetermined, and the Wigner ellipse remains a 
circle.
This is not surprising.

Now, consider the {\it non-trivial},\/ $n=-1$ case.
In such a situation, $J''/J=0$, and hence the quantity $f$ is given 
by Eq.~\eqref{eq:fnh} {\it at all times}.\/
Since $\gamma=0$, the momenta~$\cP$ defined in Eqs.~\eqref{eq:cm} 
and~\eqref{eq:cm1} turn out to be the same.
If we work with the momentum defined in Eq.~\eqref{eq:cm2}, then the 
quantity~$g$ is given by Eq.~\eqref{eq:gnh} {\it at all times}\/ so 
that the Wigner ellipse and the squeezing parameter behave as in 
the conformally invariant case.
This seems strange.
But, if we work with the conjugate momentum defined in Eq.~\eqref{eq:cm}
(as we have done in Secs.~\ref{sec:sv1}, \ref{sec:we1} and~\ref{sec:sp1}), 
then we have 
\begin{equation}
g=f'-\f{J'}{J}\,f 
\simeq -i\,\sqrt{\f{k}{2}}\, \l(1-\f{i}{k\,\eta}\r) \mathrm{e}^{-i\,k\,\eta}.
\end{equation}
This implies that the Wigner ellipse starts as a circle at early times,
while~$\mathrm{cosh}\,(2\,\sr)=1$.
However, at late times, we have  
\begin{equation}
\mathrm{cosh}\,(2\,\sr)\simeq \f{1}{2}\,\l(1+\f{\ke^2}{k^2}\r)   
\end{equation}
indicating a significant extent of squeezing for large scales.
This example confirms that our choice for the conjugate momentum $\cP$ as 
given by Eq.~\eqref{eq:cm} as an appropriate one.


\section{Details on the derivation of the entanglement entropy}\label{app:d-ee}

In this appendix, we shall briefly outline a few essential details
regarding the derivation of the entanglement entropy discussed in 
Sec.~\ref{subsec:d-ee}.

Our starting point is the definition~\eqref{eq:ee} of the entanglement 
entropy~$\mathcal{S}$ in terms of the eigen values~$p_n$ of the reduced 
density matrix~$\rho_{\mathrm{red}}(x_2,x_2',\eta)$ [cf. Eq.~\eqref{eq:rdm}].
With the aid of standard integrals (in this regard, see, for instance, 
Ref.~\cite{gradshteyn2007table}), it can be established that $p_n$ is 
given by Eq.~\eqref{eq:pn}, with $\xi$ being defined as in Eq.~\eqref{eq:xi}. 
Once we have the $p_n$ at hand, it is straightforward to sum over~$n$
in Eq.~\eqref{eq:ee} to arrive at the expression~\eqref{eq:ee-f} for 
the entanglement entropy or, equivalently, quantum discord~$\delta$.

To arrive at a result that has a simpler form when eventually expressed
in terms of the squeezing amplitude~$r$, we change to a new variable $y$, 
which is related to $\xi$ as follows:
\begin{equation}
\xi = \frac{y}{y+2}.\label{eq:def-y-app}
\end{equation}
In terms of the variable~$y$, the quantum discord~$\delta$ can be expressed 
as
\begin{eqnarray}
\delta 
&=&-\ln\, \l(\f{2}{y+2}\r) - \f{y}{2}\, \ln\l(\f{y}{y+2}\r)\nn\\
&=& \ln\, \l(1+\f{y}{2}\r) +\f{y}{2}\, \ln\l(\f{y+2}{y}\r)\nn\\
&=& \ln\, \l(\f{y+2}{2}\r) +\l(1+\f{y}{2}\r)\, 
\ln\l(\f{y+2}{y}\r)- \ln\l(\f{y+2}{y}\r)\nn\\
& =&  \l(1+\f{y}{2}\r)\, \ln\l(1+\frac{2}{y}\r)
+\ln\, \l(\f{y}{2}\r),\label{eq:ee-ff-app}
\end{eqnarray}
which is the result~\eqref{eq:ee-ff} we have quoted in the text.


\section{Another derivation of quantum discord}\label{app:ad-qd}

A convenient method to calculate the quantum discord is to write down 
the covariance matrix of the canonically conjugate variables and 
arrive at the quantum discord using the submatrices of the covariance 
matrix~\cite{adesso2010quantum,Lim:2014uea,Raveendran:2022dtb}. 
But, one has to first choose an appropriate set of two pairs of canonically 
conjugate variables, such that tracing over one set will give us the correct 
quantum discord to match with the results in the earlier literature (i.e. one
has to identify variables to represent the appropriate subsystem of the full 
system)~\cite{Lim:2014uea,Martin:2015qta}.

As we have described in Sec.~\ref{sec:i-dof}, the action describing the 
modes~$\cA_{\bmk}^{\sigma}$ of the electromagnetic field is similar to 
the action that governs the Mukhanov-Sasaki variable characterizing the 
scalar perturbations.
However, there are two differences.  
The first difference is that, due to the two states of polarization, the
electromagnetic field contains twice as many degrees of freedom as the 
scalar perturbations.
Secondly, in the helical case, due to the presence of the term that leads
to the violation of parity, the modes corresponding to the two states of 
polarization evolve differently. 
Nevertheless, the two helical states of polarizations (with $\sigma=\pm 1$) 
evolve independently, and the method adopted in the case of the scalar
perturbations can be used to characterize the quantum discord associated 
with either of the two states of the polarization of the electromagnetic 
field.

The quantum discord that we would like to calculate is when the system is
divided into modes with wave vectors $\bmk$ and $-\bmk$, as in the case of 
the scalar perturbations~\cite{Lim:2014uea,Martin:2015qta}. 
As we have explained in the main text, the correct variables to use are the 
conjugate variables $(x^\sigma_{\bm k},p_{\bm k}^{\sigma})$ as defined
in Eq.~\eqref{eq:xp-Ap}, but with $\bar{\omega}$ replaced by $\widetilde{\omega}$
[cf. Eq.~\eqref{eq:omegatilde2}].
On using our convention of referring to $x_{\bmk}^\sigma$ and $x_{-\bmk}^\sigma$ 
as~$x_1$ and~$x_2$, the covariance matrix of the two pairs of canonically 
conjugate variables $(\hat{x}_1,\hat{p}_1,\hat{x}_2,\hat{p}_2)$ has the form
\begin{widetext}
\begin{equation}
\bm{V} = 
\begin{bmatrix}
\langle \hat{x}_1^2\rangle
& \f{1}{2}\,\langle \hat{x}_1\,\hat{p}_1 + \hat{p}_1\, \hat{x}_1\rangle
& \langle \hat{x}_1\, \hat{x}_2\rangle
& \langle \hat{x}_1\,\hat{p}_2\rangle\\
\f{1}{2}\,\langle \hat{x}_1\,\hat{p}_1 + \hat{p}_1\, \hat{x}_1\rangle
& \langle \hat{p}_1^2\rangle
& \langle \hat{x}_2\,\hat{p}_1\rangle
& \langle \hat{p}_1\,\hat{p}_2\rangle\\
\langle \hat{x}_1\,\hat{x}_2\rangle  
& \langle \hat{x}_2\,\hat{p}_1\rangle
& \langle \hat{x}_2^2\rangle
& \f{1}{2}\,\langle \hat{x}_2\,\hat{p}_2 + \hat{p}_2\, \hat{x}_2\rangle\\
\langle \hat{x}_1\,\hat{p}_2\rangle
& \langle \hat{p}_1\,\hat{p}_2\rangle
& \f{1}{2}\,\langle \hat{x}_2\,\hat{p}_2 + \hat{p}_2\, \hat{x}_2\rangle
& \langle\hat{p}_2^2\rangle
\end{bmatrix}.\label{eq:cm-sfm-Vdef2}
\end{equation}
\end{widetext}
To calculate quantum discord, first we define a scaled covariance matrix 
as
\begin{equation}
\bm{\sigma}= 2\, \bm{V}.\label{sigma_def}
\end{equation}  
We next divide this $(4 \times 4)$ matrix in terms of $(2\times2)$
sub-blocks as follows:
\begin{equation}
\bm{\sigma}= \begin{bmatrix}
\bm{\alpha} &  \bm{\gamma}\\
\bm{\gamma}^T & \bm{\beta}
\end{bmatrix},\label{sig_form}
\end{equation}
where $\bm{\alpha}$, $\bm{\beta}$ and $\bm{\gamma}$ are $(2\times 2)$ matrices. 
Defining
\begin{equation} 
B=\mathrm{det.}~\bm{\beta},\label{eq:sub-det} 	
\end{equation}
the entanglement entropy for the $(x_2, p_2)$ subsystem,
say, $\mathcal{S}_2(\sigma_{\mathbbm{12}})$, can then be directly 
calculated to be (in this context, see Ref.~\cite{Serafini:2003ke})
\begin{equation}
\mathcal{S}_2(\sigma_{\mathbbm{12}}) = F(\sqrt{B}),\label{EE:cov}
\end{equation}
with the function~$F(x)$ being given by
\begin{equation}
F(x) = \l(\f{x+1}{2}\r)\, \ln\l(\f{x+1}{2}\r) 
-\l(\f{x-1}{2}\r) \ln\l(\f{x-1}{2}\r).\label{def_f}
\end{equation}	

Using the wave function~\eqref{eq:ga2} and the relations~\eqref{eq:om1-om2-omega}, 
the elements of the matrix~$\bm{\beta}$ can be evaluated to be
\begin{subequations}
\begin{eqnarray}
\langle \hat{x}^2_2\rangle
&=& \f{\Omegaar}{2\,\l(\Omegaar^2-\Omegabr^2\r)}
= \f{\vert \Omega_+ \vert^2+\widetilde{\omega}^2}{4\, \widetilde{\omega}^2\, 
\Omegapr},\\
\langle \hat{p}^2_2 \rangle 
&=&\widetilde{\omega}^2\,\langle \hat{x}_2^2 \rangle,\quad
\f{1}{2}\,\langle \hat{x}_2\, \hat{p}_2+ \hat{p}_2\, \hat{x}_2 \rangle = 0,
\end{eqnarray}
\end{subequations}
where $\Omegapr=(\Omega_++\Omega_+^\ast)/2$ represents the real part of~$\Omega_+$. 
Therefore, the determinant of $\bm{\beta}$ becomes
\begin{eqnarray}
B &=& 4\,\langle\hat{x}_2^2\rangle\,\langle \hat{p}_2^2\rangle 
- \langle \hat{x}_2\, \hat{p}_2+ \hat{p}_2\, \hat{x}_2\rangle^2\nn\\
&=& \f{\l(\vert \Omega_+ \vert^2+\widetilde{\omega}^2\r)^2}{4\, 
\widetilde{\omega}^2\, \Omegapr^2}.
\end{eqnarray}
To connect with the results in the main text, we can use the expression~\eqref{eq:y1} 
for~$y$ to obtain that  
\begin{equation}
\sqrt{B}= y+1.
\end{equation}
On substituting this expression for $B$ in Eq.~\eqref{EE:cov}, we can arrive at
the result~\eqref{eq:ee-ff} for the quantum discord we had obtained earlier.


\section{Charged scalar field under the influence of an electric field 
in a de Sitter universe}\label{app:se}

In this appendix, we shall discuss the Schwinger effect in de Sitter spacetime 
by considering the evolution of a charged scalar field in the presence of a 
constant electric field (for earlier discussions in this regard, see, for instance, 
Refs.~\cite{Frob:2014zka,Kobayashi:2014zza,Sharma:2017ivh,Banyeres:2018aax}).
We should mention that the corresponding results in flat spacetime can be arrived
at by considering the limit wherein the constant Hubble parameter in de Sitter 
vanishes.


\subsection{Equation of motion in an FLRW universe}

Consider a complex scalar field, say, $\psi$, evolving in a curved spacetime.
In the presence of an electromagnetic field described by the vector 
potential~$A_\mu$, the action governing the complex scalar field is given by
\begin{equation}
S[\psi] = -\int \d^{4}x\, \sqrt{-g}\,
\l[(D_{\mu}\psi)^\ast\,(D^{\mu}\psi) + m^{2}\,\psi\,\psi^\ast\r],
\label{eq:a-sf}
\end{equation}
where $D_\mu=(\pa_\mu-i\,e\,A_\mu)$ and $e$ denotes the electric charge.
On varying the above action, we obtain the equation of motion governing the 
scalar field to be 
\begin{equation}
\f{1}{\sqrt{-g}}\,D_\mu\l(\sqrt{-g}\,g^{\mu\nu}\,D_\nu\r)\,\psi-m^2\,\psi=0.
\label{eq:eom-phi}
\end{equation}
The strength~$E$ of the electric field can be expressed in terms of the 
field tensor $F_{\mu\nu}$ as 
\begin{equation}
F^{\mu\nu}\,F_{\mu\nu}=-2\,E^2.
\end{equation}
If we choose to work with the vector potential 
\begin{equation}
A_\mu=[0,0,0,-A(\eta)],\label{eq:Amu}
\end{equation}
then, in the FLRW universe, the electric field is oriented along the 
$z$-direction and its strength is given by
\begin{equation}
E=\f{A'}{a^2}.\label{eq:E}    
\end{equation}

If we define the new variable 
\begin{equation}
u(\eta,{\bm x}) = a(\eta)\,\psi(\eta,{\bm x}),\label{eq:u}
\end{equation}
then, for the FLRW line-element~\eqref{eq:FLRW} and the vector 
potential~\eqref{eq:Amu}, the action~\eqref{eq:a-sf} takes the form
\begin{eqnarray}
S[u] &=& \int \d\eta\, \int \d^{3}\bm{x}\, \biggl\{\vert u'\vert^2
-\vert{\bm \pa}_{\perp} u\vert^2-\vert D_{z} u\vert^2\nn\\
& &-\,\f{a'}{a}\,\l(u\,u'{}^\ast+ u^\ast\, u'\r)
- \l[m^2\, a^2 -\l(\f{a'}{a}\r)^2\r]\,\vert u\vert^2\biggr\},\nn\\
\end{eqnarray}
where ${\bm \pa}_\perp=(\pa_x,\pa_y)$ and $D_z=\pa_z + i\, e\, A(\eta)$.
The symmetries of the FLRW metric and the fact that the vector 
potential~$A_\mu$ depends only on the conformal time coordinate 
allows us to decompose the quantity~$u(\eta,{\bm x})$ as follows:
\begin{equation}
u(\eta,{\bm x})=\int\f{\d^3{\bmk}}{(2\,\pi)^{3/2}}\,q_{\bmk}(\eta)\;
\mathrm{e}^{i\,{\bmk}\cdot{\bm x}}.\label{eq:fd}
\end{equation}
We should point out that, since~$u$ is a complex field, we do {\it not}\/ 
have a condition connecting the Fourier modes~$q_{\bmk}$ and~$q_{-\bmk}$ 
akin to~Eq.~\eqref{eq:rcf}. 
The action in Fourier space that governs the modes $q_{\bmk}$ can be
obtained to be
\begin{eqnarray}
S[q_{\bmk}] 
&=& \int \d\eta\,\int \d^{3}\bmk\,
\biggl\{\vert q_{\bmk}'\vert^2
-\f{a'}{a}\,\l(q_{\bmk}\,q_{\bmk}'{}\!^\ast+ q_{\bmk}'\, q_{\bmk}^\ast \r)\nn\\
& &-\, \mu_q^2\, \vert q_{\bmk}\vert^2 \biggr\},\label{eq:S1}
\end{eqnarray}
where the quantity $\mu_q^2$ is given by
\begin{equation}
\mu_q^2(\eta) 
= k_\perp^2 + (k_z+e\,A)^2+m^2\, a^2-\l(\f{a'}{a}\r)^2
\end{equation}
with ${\bmk}_\perp=(k_x,k_y)$ and $k_\perp=\vert {\bmk}_\perp\vert$.
Thus, each mode with wave vector~$\bmk$ evolves independently according to 
identical (though $\bmk$-dependent) actions.

Let us now express $q_{\bmk}$ as
\begin{equation}
q_{\bmk}=\f{1}{\sqrt{2}}\,\l(q_{\bmk{{\mathrm{R}}}}
+i\,q_{\bmk\mathrm{I}}\r),
\end{equation}
where $q_{{\bmk\mathrm{R}}}$ and $q_{{\bmk\mathrm{I}}}$ are the real
and imaginary parts of $q_\bmk$.
The actions for $q_{\bmk\mathrm{R}}$ and $q_{\bmk\mathrm{I}}$ decouple
and are identical in form.  
Therefore, the dynamics of the system can be analyzed using the 
following Lagrangian density in Fourier space:
\begin{equation}
\mathcal{L}=\f{1}{2}\,q'{}^2 -\f{a'}{a}\,q\,q' 
-\f{1}{2}\,\mu_q^2\, q^2,\label{eq:Lq}
\end{equation}
where $q$ stands for either~$q_{{\bmk\mathrm{R}}}$ or~$q_{{\bmk\mathrm{I}}}$.
The momentum conjugate to the variable $q$ is given by
\begin{align}
p_{q} = q' -\f{a'}{a}\, q.  \label{eq:dp}
\end{align}
The Lagrangian~\eqref{eq:Lq} leads to following equation of motion for the
Fourier mode~$q$:
\begin{equation}
q''+\omega_q^2\,q=0,\label{eq:q-eom}
\end{equation}
where the quantity $\omega_q^2$ is given by
\begin{equation}
\omega_q^2(\eta) = k_\perp^2 + (k_z+e\,A)^2+m^2\, a^2-\f{a''}{a}.
\end{equation}


\subsection{Solutions in de Sitter and the behavior of the squeezing amplitude}

Let us now discuss the solutions to the modes of the scalar field in 
the presence of a constant electric field in de Sitter spacetime and 
the behavior of the corresponding squeezing
amplitude~\cite{Frob:2014zka,Sharma:2017ivh}.
Earlier, in Sec.~\ref{sec:dsi}, we had assumed that the scale factor in 
de Sitter inflation is given by $a(\eta)=-1/(\HI\,\eta)$, where $\HI$
is a constant.
Instead, we shall now assume that the de Sitter spacetime is described by 
the scale factor
\begin{equation}
a(\eta)=\f{1}{1-\HI\,\eta},
\end{equation}
where~$-\infty<\eta<\HI^{-1}$. 
We have chosen such a form since we can obtain the Minkowski spacetime 
as the limit $\HI\to0$.
Since the strength of the electric field is given by $E=A'/a^2$ 
[see Eq.~\eqref{eq:E}], if we require $E$ to be constant, then for the 
above choice of the scale factor, the vector potential $A(\eta)$ 
is given by
\begin{equation}
A(\eta)=\f{E_0}{\HI}\,\l[a(\eta)-1\r],    
\end{equation}
where $E_0$ is a constant and we have chosen the constant of integration 
to be $-E_0/\HI$.
Such a choice allows us to have a well behaved $\HI\to 0$ limit of
$A(\eta)$, which reduces to the usual choice of vector potential 
considered to examine the Schwinger effect in Minkowski spacetime.

Evidently, we can quantize the system described by the Lagrangian~\eqref{eq:Lq}
in the same manner as we quantized the electromagnetic vector 
potential~$\cA$ in the \Sch picture in Sec.~\ref{sec:sv1}.
For the above choices of the scale factor and the vector potential, we find that 
the function~$f$ that determines the wave function describing the system [given by
Eqs.~\eqref{eq:wfncA} and~\eqref{eq:domega}, and $g$ defined as in Eq.~\eqref{eq:g}, 
with the non-conformal coupling function~$J$ replaced by the scale factor~$a$] 
satisfies the differential equation
\begin{equation}
\f{\d^2 f}{\d \tau^2}
+\l(\bar{k}^2-\f{2\,\zeta\,\bar{k}_z}{\tau}
+\f{\zeta^2+\bar{m}^2-2}{\tau^2}\r)\, f=0, \label{eq:q}
\end{equation}
where $\bar{k}_z=k_z-(e\,E_0/\HI)$, $\bar{k}^2=k_\perp^2+\bar{k}_z^2$,
$\zeta=e\,E_0/\HI^2$, $\bar{m}=m/\HI$ and $\tau=\eta-(1/\HI)$.
We should point out here the similarity between the above differential equation 
and the equation~\eqref{eq:cA-h-de-sp} governing the evolution of the helical
electromagnetic fields in de Sitter spacetime.
Note that, for small~$\bar{m}$ and~$\zeta$, their structure are very similar. 
Also, for a range of wave numbers, changing the sign of $k_z$ (or the direction 
of the electric field~$E$) is equivalent to considering the helical electromagnetic 
mode of opposite polarization.

\begin{figure}[t]
\centering
\includegraphics[width=1.0\linewidth]{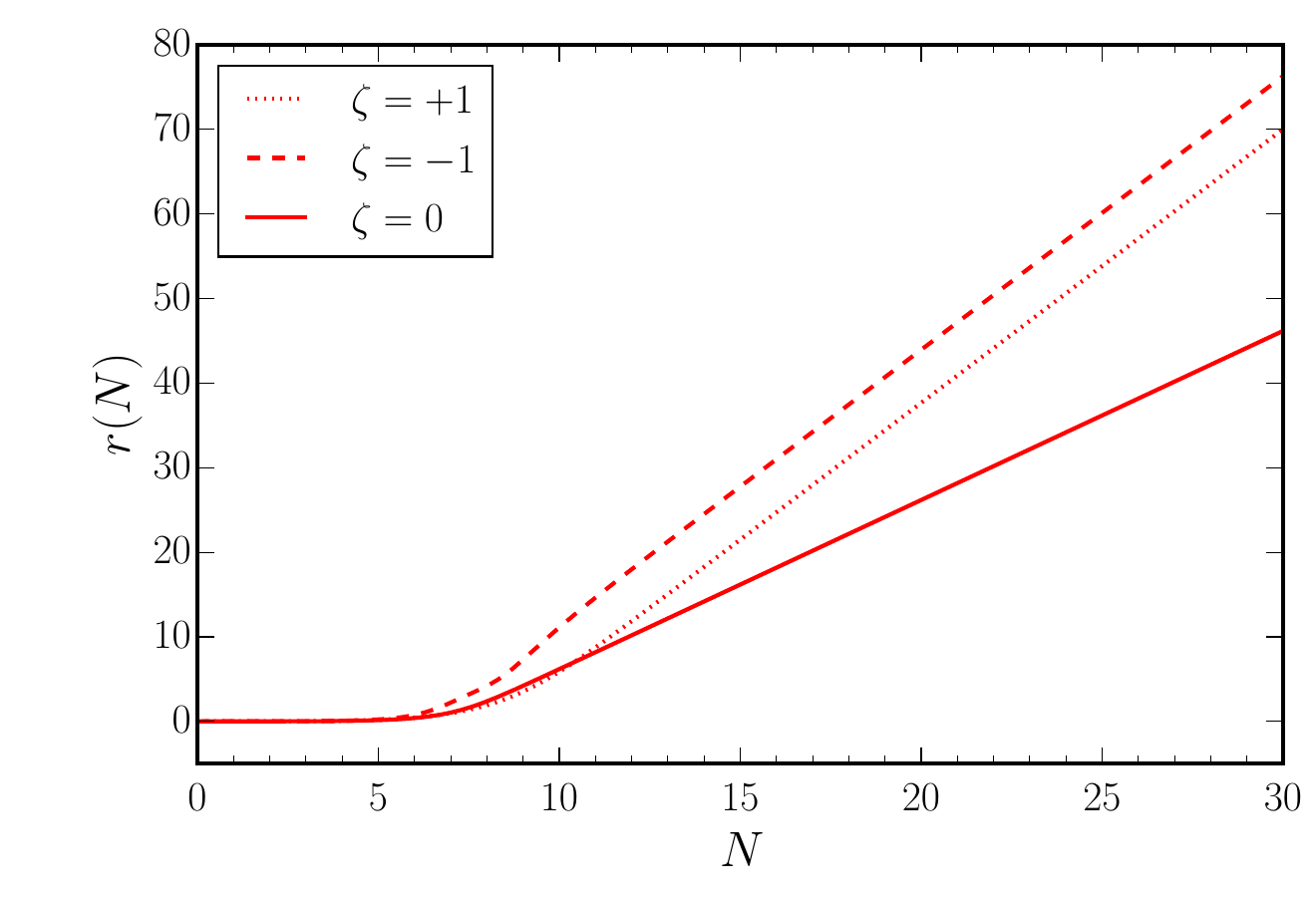}
\vskip -15pt
\caption{The behavior of the squeezing parameter~$r(N)$ for the case 
of a charged scalar field in de Sitter inflation has been plotted as 
a function of $e$-folds $N$ for the following choices of the 
parameters:~$(k_\perp,\bar{k}_z)=(0,10^{-2})$, $\bar{m}=0$ and 
$\zeta=(-1,0,1)$ (in dashed, solid and dotted red).
Note that, in a manner similar to that of the $\sigma=-1$ helical
electromagnetic mode, the squeezing amplitude $r$ is larger when 
$\zeta=-1$ than the case wherein $\zeta=+1$.  
For the wave number we have worked with, we find that the modes can 
be said to leave Hubble radius around $N\simeq 7$.}\label{fig:r-csf}
\end{figure}
Recall that, if the wave function describing the mode is to start from the 
ground state corresponding to the Bunch-Davies vacuum, then, as $(-k\,\eta)\to \infty$,
we require that the function $f$ behaves as $f \propto \mathrm{exp}\,(-i\,\bar{k}\, 
\eta)$.
It is straightforward to show that the modes with such initial conditions 
are given by 
\begin{eqnarray}
f(\tau) &=& \f{1}{\sqrt{2\,\bar{k}}}\,\mathrm{e}^{-\pi\, \zeta\,\bar{k}_z/(2\,\bar{k})}\,
\mathrm{e}^{-i\,\bar{k}/\HI}\, 
W_{i\,\zeta\,\bar{k}_z/\bar{k},\nu}(2\,i\,\bar{k}\,\tau),\label{eq:m-ds}\nn\\
\end{eqnarray}
where $W_{\lambda,\nu}(z)$ denotes the Whittaker function and $\nu^2=(9/4)
-\bar{m}^2-\zeta^2$~\cite{gradshteyn2007table}.
Note that, when $\zeta=0$ and $m=0$, the above solution reduces to 
\begin{equation}
f(\eta) 
=\f{1}{\sqrt{2\,k}}\, \mathrm{e}^{-i\,k/\HI}\, 
W_{0,3/2}(2\,i\,k\,\tau).
\end{equation}
Since~\cite{gradshteyn2007table}
\begin{equation}
W_{0,3/2}(z)=\sqrt{\f{z}{\pi}}\,K_{3/2}\l(\f{z}{2}\r),
\end{equation}
where $K_{3/2}(z)$ is the modified Bessel's function given by
\begin{equation}
K_{3/2}(z)=\sqrt{\f{\pi}{2\,z}}\,\l(1+\f{1}{z}\r)\,\mathrm{e}^{-z},    
\end{equation}
we find that the above function~$f(\eta)$ can be expressed as
\begin{equation}
f(\eta) =\f{1}{\sqrt{2\,k}}\,
\l[1-\f{i}{k\,\eta-(k/\HI)}\r]\,
\mathrm{e}^{-i\,k\,\eta},
\end{equation}
which is the well known solution describing a massless scalar field
in de Sitter spacetime.

The squeezing amplitude associated with the modes of the charged scalar 
field can be determined using the relation~\eqref{eq:r-g}, with $f$ given 
by Eq.~\eqref{eq:m-ds} and $g$ defined as in Eq.~\eqref{eq:g}, with $J$ 
replaced by~$a$. 
In Fig.~\ref{fig:r-csf}, we have plotted the evolution of the squeezing 
amplitude $\sr$ (for a specific wave number) as a function of $e$-folds 
for two different sets of values of the parameters~$\zeta$, with $\bar{m}$
set to zero.
We have also plotted for the case wherein $\zeta$ vanishes.
It should be evident that the behavior of the squeezing amplitude for 
opposite signs of~$\zeta$ is similar to the behavior of the two states 
of opposite polarization of the helical electromagnetic mode.
In fact, because of the presence of the additional parameters (such as~$m$),
the evolution of the complex scalar field is considerably richer, but we 
have chosen to work with values so that the evolution closely resembles the
behavior of the modes of the non-helical and helical electromagnetic fields.  

\bibliographystyle{apsrev4-2}
\bibliography{references,mybibliography-cosmology,mybibliography-quantum,mybibliography-maths,mybibliography-qft} 

\end{document}